\documentclass[twocolumn]{aa} %
\usepackage[OT1]{fontenc}
\usepackage{graphicx}
\usepackage{epsfig}
\usepackage{natbib}
\bibpunct{(}{)}{;}{a}{}{,} 

\def\mathrelfun#1#2{\lower3.6pt\vbox{\baselineskip0pt\lineskip.9pt
  \ialign{$\mathsurround=0pt#1\hfil##\hfil$\crcr#2\crcr\sim\crcr}}}

\def\lesssim{\mathrel{\mathpalette\fun <}} 

\def\fun#1#2{\lower3.6pt\vbox{\baselineskip0pt\lineskip.9pt
  \ialign{$\mathsurround=0pt#1\hfil##\hfil$\crcr#2\crcr\sim\crcr}}}

\begin{document}

\title{Constraints on the non-linear coupling parameter $f_{nl}$ with Archeops data}

\author{ 
A.~Curto ~\inst{1} ~\inst{2} \and
J.~F.~Mac\'{\i}as--P\'erez ~\inst{3} \and  
E.~Mart\'{\i}nez-Gonz\'alez ~\inst{1} \and
R.~B.~Barreiro ~\inst{1} \and
D.~Santos~\inst{3} \and
F. K.~Hansen~\inst{4} \and
M. ~Liguori ~\inst{5} \and
S. ~Matarrese ~\inst{6}
}

   \offprints{curto@ifca.unican.es}

   \institute{
     Instituto de F\'isica de Cantabria, CSIC-Universidad de Cantabria, Avda. de los Castros s/n, 39005 Santander, Spain
    \and
    Dpto. de F\'isica Moderna, Universidad de Cantabria, Avda. los Castros s/n, 39005 Santander, Spain
    \and
	LPSC, Universit\'e Joseph Fourier Grenoble 1, CNRS/IN2P3, Institut National Polytechnique de Grenoble, 53, av. des Martyrs, 38026 Grenoble, France 
    \and
    Institute of Theoretical Astrophysics, University of Oslo, P.O. Box 1029 Blindern, 0315 Oslo, Norway
    \and
    Department of Applied Mathematics and Theoretical Physics, Centre for Mathematical Sciences, University of Cambridge, Wilberfoce Road, Cambridge, CB3 0WA, United Kingdom
    \and
    Dipartimento di Fisica “G. Galilei”, Universit\`a di Padova and INFN, Sezione di Padova, via Marzolo 8, I-35131, Padova, Italy
}
   
  \date{\today}
  
  \abstract{} { We present a Gaussianity analysis of the Archeops
    Cosmic Microwave Background (CMB) temperature anisotropy data maps at high
    resolution to constrain the non-linear coupling parameter
    $f_{nl}$ characterising well motivated non-Gaussian CMB models.
    We used the data collected by the most sensitive Archeops
    bolometer at 143~GHz.
    The instrumental noise was carefully
    characterised for this bolometer, and for another Archeops
    bolometer at 143~GHz used for comparison. Angular scales from 27
    arcmin to 1.8 degrees and a large fraction of the sky, 21\%,
    covering both hemispheres (avoiding pixels with Galactic latitude
    $|b| < 15$ degrees) were considered.
     }
    { The three Minkowski functionals on the sphere evaluated at different
      thresholds were used to construct a $\chi^2$ statistics for both
      the Gaussian and the non-Gaussian CMB models.  The
      Archeops maps were produced with the Mirage optimal map-making
      procedure from processed time ordered data.  The analysis is
      based on simulations of signal (Gaussian and non-Gaussian
      $f_{nl}$ CMB models) and noise which were processed in the time
      domain using the Archeops pipeline and projected on the sky
      using the Mirage optimal map-making procedure.
    }
    { The Archeops data were found to be compatible with Gaussianity 
      after removal of highly noisy pixels at high resolution. The
      non-linear coupling parameter was constrained to $f_{nl}=
      70_{-400}^{+590}$ at 68\% CL and $f_{nl}=70_{-920}^{+1075}$ at
      95\% CL, for realistic non-Gaussian CMB simulations.
    } {}


\maketitle


\section{Introduction}
The Cosmic Microwave Background (CMB) radiation is a valuable tool to
study the early universe. When we observe this weak radiation we are
observing the universe when it was about 400000 years old. Several
theories have been proposed to explain the origin and the evolution of
the universe. The Big Bang is the most favored theory and it predicts
the existence of the CMB. The inflationary models are complementary to
the Big Bang model. Among their many properties we are interested in
those related to the CMB and the predictions that they present about
it. The standard, single field, slow roll inflation
\citep{guth,albrecht,linde1982,linde1983} is one of the most accepted
models because of the accuracy of its predictions and the observations
performed with modern experiments \citep{spergel}. In particular, the
standard inflation predicts that the primordial density fluctuations
are distributed following a nearly Gaussian distribution. These
fluctuations are imprinted in the anisotropies of the CMB. Therefore
if the prediction is correct the anisotropies of the CMB should be
distributed in a nearly Gaussian way. Any detection of non-Gaussian
deviations would have far-reaching consequences for our understanding
of the Universe \citep{cruzc,wandelt}. In addition the search for
non-Gaussian fluctuations in the data is a useful tool to look for
residual foreground, secondary anisotropies or unidentified systematic
errors. Some of these effects may introduce non-Gaussian features at
different levels.

Many statistical tools have been applied to test the Gaussianity of
CMB data sets. The Minkowski functionals have been applied to
different recent experiments \citep{spergel,troia,gott2007,curto_32};
other examples are the smooth tests of goodness-of-fit
\citep{cayon,aliaga_vsa,rubino,curto_32}, wavelets
\citep{barreiro2000,cayon2001,vielva2004,mukherjee2004,cruza,cruzb},
local estimators of the n-point correlations
\citep{eriksen2004,eriksen2005}, steerable filters
\citep{wiaux,vielva2007}, and the CMB 1-pdf \citep{monteser2007} among
others.

In this work we analyse the CMB data collected by the Archeops
experiment. Important results have been obtained from this experiment
since its launch in 2002. It gave the first link in the $C_\ell$
determination \citep{archpaper} between the COBE large angular scale
data \citep{cobe} and the first acoustic peak as measured by BOOMERanG
and MAXIMA \citep{boomerang,maxima}. From this it gave a precise
determination of the main cosmological parameters
\citep{archpaper_cospar}. It also provided an independent confirmation
at different frequencies of the power spectrum for the range $\ell =
10$ to $\ell = 700$ of the WMAP first year results
\citep{tristram_cl}.

In this study we use the Minkowski functionals
\citep{minkowski1903,gott1990,schmalzing}.  This new Gaussianity
analysis of the Archeops data complements the first analysis presented
in \citet{curto_32} where only low resolution maps (about 1.8 degrees
of resolution) were analysed and the $f_{nl}$ constraints were
imposed on non-Gaussian CMB maps simulated using the quadratic
Sachs-Wolfe approximation.  In contrast, in the present work, the
constraints on the non-linear coupling parameter $f_{nl}$ are obtained
using higher resolution (27 arcmin) Archeops data and realistic
non-Gaussian simulations of the CMB fluctuations with the algorithms
developed by \citet{liguori2003,liguori2007}.

Our article is organized as follows. Section \ref{datamethod} presents
the Archeops data and the Gaussian and non-Gaussian simulations
performed.  Section 3 describes the statistical methods to test
Gaussianity and to constrain the $f_{nl}$ parameter. In Section 4 we
perform an analysis of the instrumental noise which will provide the
correct masks for our analysis. Section 5 is devoted to the $f_{nl}$
constraints and the comparison with the Sachs-Wolfe approximation. We
summarize and draw our conclusions in Section 6.
%
%
\section{Data and simulations}
\label{datamethod}
\subsection{The data}
Archeops\footnote{http://www.archeops.org} is a balloon-borne
experiment to map the CMB anisotropies at high resolution with a large
sky coverage. The experiment is described in
\citet{trapani,archeopsballona,archeopsballon,archeops_tecnical}.  It
is based on the Planck high frequency instrument (HFI) technology. The
instrument consists of a 1.5 m telescope pointing at 49 degrees from
its vertical axis on board a stratospheric gondola. It has 21
bolometers cooled to 100 mK by an $^3$He/$^4$He dilution cryostat
designed to work at similar conditions to the ones expected for
Planck. These bolometers operate at frequencies of 143, 217, 353 and
545 GHz. After one test flight in Trapani (Italy), the instrument flew
three times from the CNES/Swedish facility of Esrange, near Kiruna
(Sweden) to Siberia (Russia). The last flight on February $7^{th}$
2002 provided 12.5h of CMB-quality data for a total of 19h. The
experiment was performed during the Arctic winter to avoid Sun
contamination. These data correspond to a sky coverage of
approximately 30\% of the sky, including the Galactic plane. From its
four frequency bands the two lowest (143 and 217 GHz) were dedicated
to the observation of the CMB and the others (353 and 545 GHz) to the
monitoring and calibration of both atmospheric and Galactic emission.

We used the data collected with two bolometers at 143 GHz. We used the
data of the $143K03$ bolometer for the main analysis (see
Sect.~\ref{results}) and the data of the $143K03$ and $143K04$
bolometers for the noise analysis (see
Sect.~\ref{noiseanalysis}). However, for the $f_{nl}$ constraints, the
data from the second bolometer $143K04$ are not used due to their
higher noise level and worse systematic errors.  The characteristics
of the bolometers are described in \citet{archeops_tecnical}.  In
\citet{curto_32} the Gaussianity analysis of Archeops data was
performed at low resolution, in particular HEALPix \citep{healpix}
$N_{side}=$ 32. Here we complement that work and analyse the data at
the same and higher resolutions: $N_{side}=$ 32, 64, and 128 using the
realistic non-Gaussian simulations presented below.

First we processed and cleaned the Time Ordered Information (hereafter
TOI) as described in \citet{tristram_cl}.  Then the data maps at
different resolutions were produced using the Mirage optimal
map-making procedure \citep{mirage}.  All the analyses presented here
were performed on a fraction of the Archeops observed region after
masking out pixels with Galactic latitude $|b| < 15^o$. This
corresponds to 21\% of the total sky.  Unlike the analysis in
\citet{curto_32}, restricted to north Galactic latitudes, south
Galactic latitudes are included in the analysis.
\subsection{Gaussian simulations}
We have performed Gaussian simulations of the CMB signal and of the
Archeops noise as described in \citet{curto_32}. The noise simulations
were obtained from Gaussian random realisations of the time-domain
noise power spectrum of the bolometers. The constructed noise time
ordered data were then projected using the Mirage optimal map-making
procedure.  The CMB Gaussian simulations were obtained from random
realisations of the best--fit Archeops CMB power spectrum. From these
maps we constructed Archeops time ordered data and then we project
them using Mirage. In this way, the filtering of the Archeops data was
taken into account.
\subsection{Non-Gaussian CMB simulations}
\label{model}
There are several alternative theories to the standard inflation
theory that introduce non-Gaussian fluctuations in the CMB.  One
simple model that describes weakly non-Gaussian fluctuations in matter
and radiation is obtained by introducing a quadratic term in the
primordial gravitational potential
\citep{salopek,gangui,verde,komatsu2001}
\begin{equation}
\Phi(\vec x) = \Phi_L(\vec x) + f_{nl}\{\Phi_L^2(\vec x)-\langle \Phi_L^2(\vec x) \rangle \} \ \ 
\end{equation}
where $\Phi_L(\vec x)$ is a linear random field which is Gaussian
distributed and has zero mean, and $f_{nl}$ is is the non-linear
coupling parameter. Non-Gaussianity of this type is generated in
various classes of non-standard inflationary models \citep[see,
  e. g.][for a review]{bartolo}. To obtain the CMB anisotropies
generated by such a primordial gravitational potential we use the
algorithm described in \citet{liguori2003} for the temperature maps or
\citet{liguori2007} for temperature and polarization maps.

In this case the multipole coefficients $a_{lm}$ of the CMB
temperature map can be written as
\begin{equation}
a_{lm} = a_{lm}^{(G)}+f_{nl}a_{lm}^{(NG)}
\end{equation}
where $a_{lm}^{(G)}$ is the Gaussian contribution and $a_{lm}^{(NG)}$
is the non-Gaussian contribution.

For our simulations we use the $\Lambda$CDM model that best fits the
WMAP data and a modified version of the CMBFAST code \citep{cmbfast}
to obtain the Gaussian and non-Gaussian contributions as described
above.  We produced a set of 300 high resolution full sky temperature
Gaussian maps $\Delta T_{G}$ and their complementary non-Gaussian maps
$\Delta T_{NG}$ at the same resolution. The total CMB map with a
non-Gaussian contribution is therefore
\begin{equation}
\Delta T = \Delta T_{G} +f_{nl} \Delta T_{NG}.
\label{model1}
\end{equation}
To constrain $f_{nl}$ with Archeops data we need to transform these
simulations into Archeops simulations. For this purpose, each
simulated non-Gaussian map given by Eq. \ref{model1} was converted
into Archeops time-ordered data accounting for the Archeops pointing.
These time-ordered data were re-normalised to account for
intercalibration errors between Archeops and WMAP\footnote{The
  normalisation is simply obtained by multiplying the simulated TOI by
  a constant factor $f = 1/1.07$ as described
  in~\citet{tristram_cl}.}, and added to Gaussian simulations of the
Archeops instrumental noise constructed as above.  Then the time
ordered data was projected on the sky using Mirage.  Although, the
Mirage algorithm is weakly non-linear, i.e., the map obtained from
$\Delta T_G$ and $\Delta T_{NG}$ separately is slightly different from
the one obtained from the sum of the two contributions, we have tested
that the Minkowski functionals are very similar in both cases. Thus,
for our analysis, we can construct the Archeops simulations for
different values of $f_{nl}$ by adding the Mirage map for $\Delta T_G$
to the one for $\Delta T_{NG}$ multiplied by $f_{nl}$.  This
corresponds to an important saving of CPU time as the Mirage algorithm
is a significantly time consuming process \citep[][]{curto_32}.

Thus, we transformed our sets of 300 $\Delta T_{G}$ and $\Delta
T_{NG}$ simulations into CMB Archeops simulations ($s_g$ and $s_{ng}$)
at the three considered resolutions $N_{side}=$32, 64, and 128. The
final non-Gaussian simulations accounting for $f_{nl}$ were computed
from $s_{g}+f_{nl}s_{ng}+n_g$ where $n_g$ corresponds to the Archeops
Gaussian noise simulations.
\section{Methodology}
\label{methodology}
The Gaussianity analysis and the constraints on the $f_{nl}$ parameter
were performed using the Minkowski functionals. Detailed theoretical
information about these quantities is presented for example in
\citet{gott1990,schmalzing}. Some recent examples of applications of
the Minkowski functionals to the CMB are in \citet{curto_32} for
Archeops data at low angular resolution, \citet{troia} for BOOMERanG
2003 data, and \citet{spergel} for WMAP 3rd year data.

For a scalar field $\Delta T (\vec n)$ in the sphere we have three
Minkowski functionals given a threshold $\nu$. Considering the
excursion set of points $Q_{\nu}$ where $\Delta T ( \vec n)/ \sigma >
\nu$ the three Minkowski functionals are: the area $A( \nu )$ of
$Q_\nu$, the contour length $C(\nu)$ of $Q_\nu$, and the genus
$G(\nu)$ (proportional to the difference between hot spots above $\nu$
and cold spots below $\nu$). The expected values of the functionals
for a Gaussian random field are \citep{schmalzing}
\begin{eqnarray}
\nonumber
\langle A(\nu) \rangle &   = &  \frac{1}{2} \left(1  - \frac{2}{\sqrt{\pi}}\int_{0}^{\nu/\sqrt{2}}exp(-t^2)dt\right) \\
\label{area}
\nonumber
\langle C(\nu) \rangle & = &  \frac{\sqrt{\tau}}{8}exp\left(-\frac{\nu^2}{2}\right)  \\
\label{contour}
\nonumber
\langle G(\nu) \rangle & =  & \frac{\tau}{(2\pi)^{3/2}} ~ \nu ~ exp\left( -\frac{\nu^2}{2}\right)\\
\label{genus}
\end{eqnarray}
where $\sigma = \sum_{\ell=1}^{\ell_{max}}(2\ell+1)C_\ell$ and $\tau =
\sum_{\ell=1}^{\ell_{max}}(2\ell+1)C_\ell\ell(\ell+1)/2$. These
expressions are only valid in an ideal case.  In practice, we will
obtain these quantities from Gaussian CMB simulations in order to take into
account the noise, mask and pixel effects.

For a Gaussian random field the Minkowski functionals are
approximately Gaussian distributed, therefore we can use a $\chi^2$
statistic to test the Gaussianity of a CMB map. This statistic is
computed with the three Minkowski functionals evaluated at $n_{th}$
different thresholds.
\begin{equation}
\chi^2=\sum_{i,j=1}^{3n_{th}}(v_i-\langle v_i \rangle)C_{ij}^{-1}(v_j-\langle v_j \rangle)
\label{chi2}
\end{equation}
where $\langle \rangle$ is the expected value for the Gaussian case,
$\vec v$ is a $3 n_{th}$ vector 
\begin{equation}
\label{vectorv}
(A(\nu_1),..,A(\nu_{n_{th}}),C(\nu_1),..,C(\nu_{n_{th}}),G(\nu_1),..,G(\nu_{n_{th}})),
\end{equation}
and $C$ is a $3 n_{th}$ covariance matrix 
\begin{equation}
\label{correl}
C_{ij} = \langle v_i v_j \rangle - \langle v_i \rangle \langle v_j \rangle.
\end{equation}
The Gaussianity analysis consists of computing the $\chi^2$ statistic
of the data and compare it with the value of the $\chi^2$ statistic of
a set of Gaussian simulations of the data (CMB signal plus
instrumental noise). 

Another important analysis is the estimation of the $f_{nl}$
parameter. In this case, we can use a $\chi^2$ test with the Minkowski
functionals
\begin{equation}
\chi^2(f_{nl})=\sum_{i,j=1}^{3n_{th}}(v_i-\langle v_i \rangle_{f_{nl}})C_{ij}^{-1}(f_{nl})(v_j-\langle v_j \rangle_{f_{nl}})
\end{equation}
where $\langle \rangle_{f_{nl}} $ is the expected value for a model
with $f_{nl}$, and $C_{ij}(f_{nl}) = \langle v_i v_j \rangle_{f_{nl}}
- \langle v_i \rangle_{f_{nl}} \langle v_j \rangle_{f_{nl}}$. For low
values of $f_{nl}$ ($f_{nl} \lesssim 1500$) we have that
$C_{ij}(f_{nl}) \simeq C_{ij}(f_{nl}=0) = C_{ij}$, and therefore we
can use the approximation
\begin{equation}
\chi^2(f_{nl})=\sum_{i,j=1}^{3n_{th}}(v_i-\langle v_i \rangle_{f_{nl}})C_{ij}^{-1}(v_j-\langle v_j \rangle_{f_{nl}}).
\label{chifnl}
\end{equation}
The best-fit $f_{nl}$ for the Archeops data is obtained
by minimization of $\chi^2(f_{nl})$. Error bars for it
at diffirent confidence levels are computed using the
Gaussian simulations.

Other than the analysis of the data at each resolution separately we
can also analyse the data by combining the information at different
resolutions.  Assuming maps at $n_{res}$ different resolutions we can
define a vector
\begin{equation}
\vec V=(\vec v_1,\vec v_2, ..., \vec v_{n_{res}}),
\label{bigvect}
\end{equation}
where each $v_i$ corresponds to the vector given by Eq. \ref{vectorv}
for the resolution $i$. With this combined vector we can compute
$\chi^2$ statistics in the same way as in Eqs. \ref{chi2} and
\ref{chifnl}
\begin{eqnarray}
\label{bigchi2}
\chi^2 & = & \sum_{i,j=1}^{3N_{th}}(V_i-\langle V_i \rangle)C_{ij}^{-1}(V_j-\langle V_j \rangle)\\
\label{bigchifnl}
\chi^2(f_{nl}) & = & \sum_{i,j=1}^{3N_{th}}(V_i-\langle V_i \rangle_{f_{nl}})C_{ij}^{-1}(V_j-\langle V_j \rangle_{f_{nl}})
\end{eqnarray}
where $\langle \rangle$ is the expected value for the Gaussian case,
$\langle \rangle_{f_{nl}} $ is the expected value for a model with
$f_{nl}$, $N_{th}=\sum_{k=1}^{n_{res}}n_{th}^{(k)}$, $n_{th}^{(k)}$
is the number of thresholds used at resolution $k$, and
$C_{ij} = \langle V_i V_j \rangle - \langle V_i \rangle \langle V_j
\rangle$.
%
%
\section{Gaussianity analysis of the Archeops instrumental noise}
\label{noiseanalysis}
\begin{figure*}[t]
\center
\epsfig{file=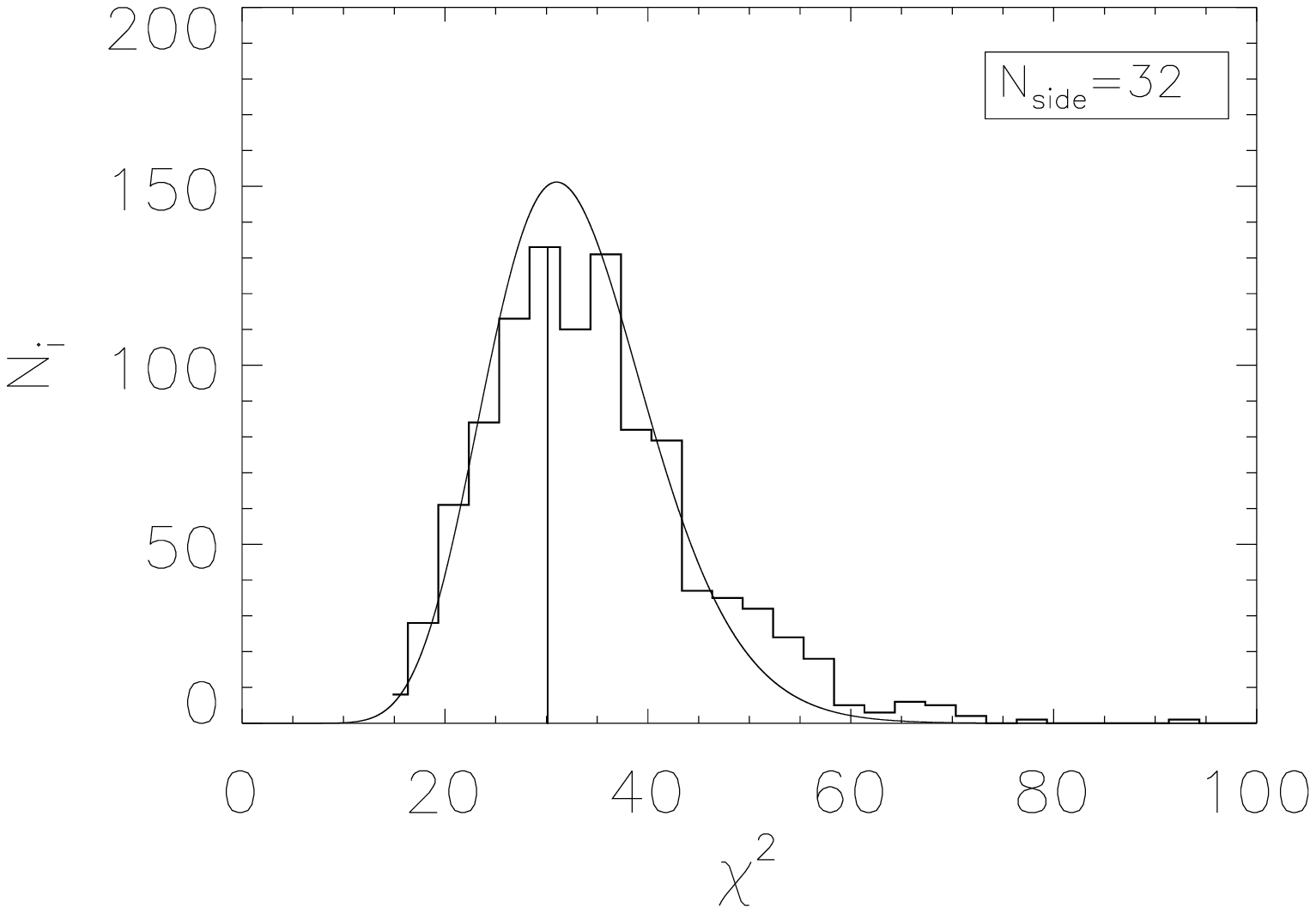,height=3.7cm,width=5.9cm}
\epsfig{file=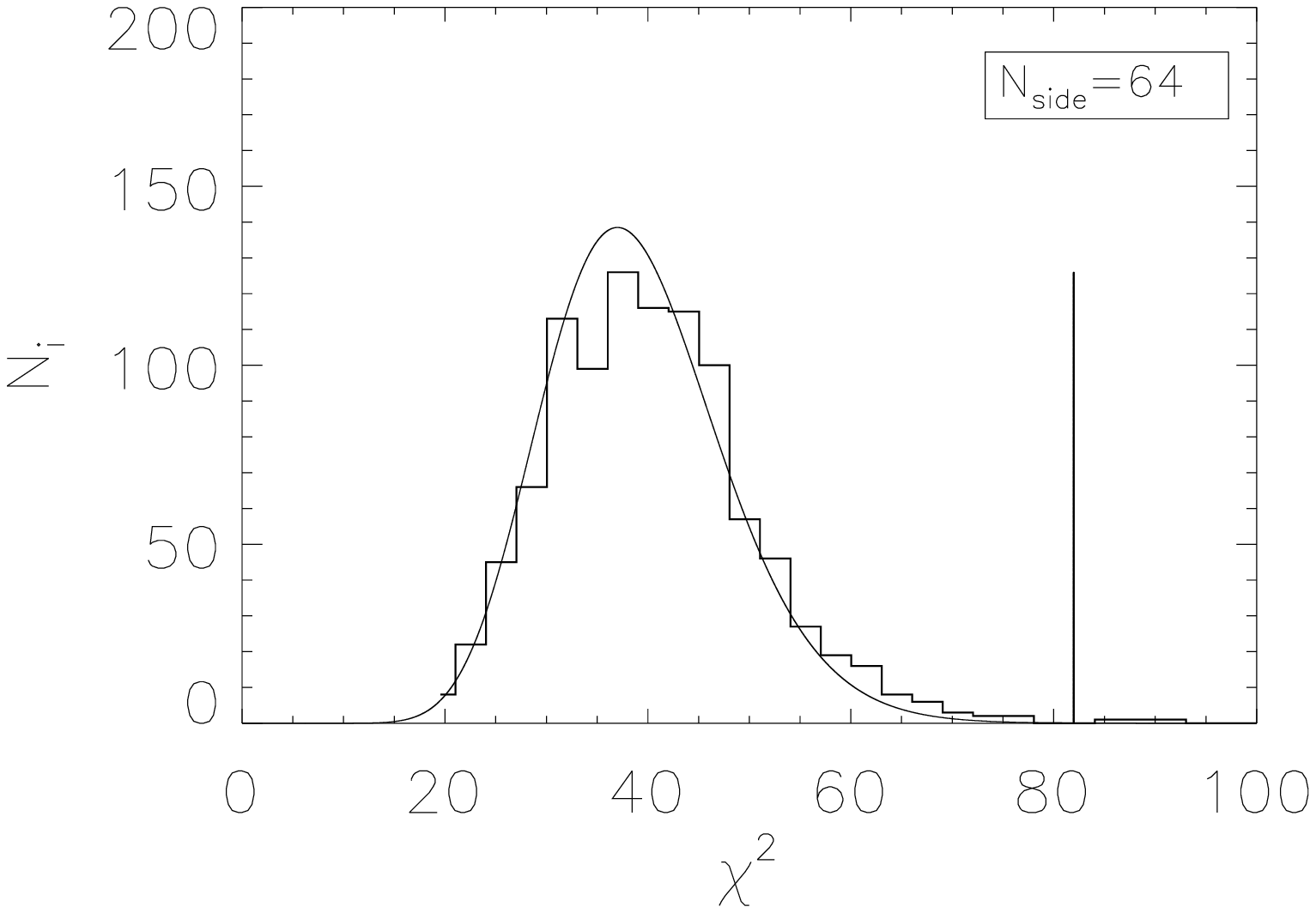,height=3.7cm,width=5.9cm}
\epsfig{file=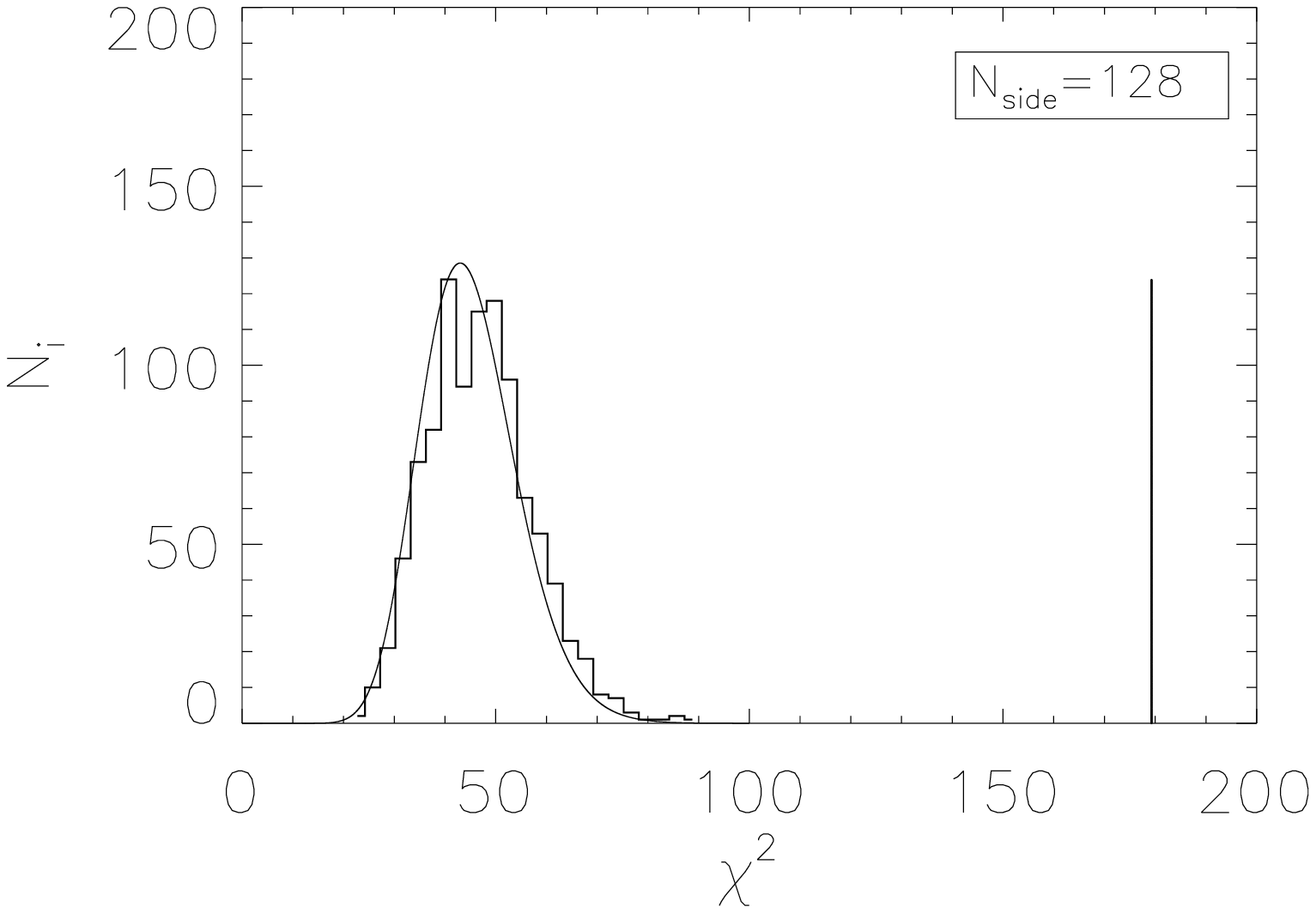,height=3.7cm,width=5.9cm}
\caption{
  Distribution of the $\chi^2$ values
  from the Minkowski Gaussianity analysis for the Archeops 143K03-143K04
  map at $N_{side}=32$, $N_{side}=64$, and $N_{side}=128$. The vertical
  lines show data, the histograms are obtained from a set of
  $10^3$ simulations, and the solid lines are the expected $\chi^2$
  distributions for $3n_{th}$ degrees of freedom.}
\label{d1minusd2gauss}
\end{figure*}
\begin{figure*}[t]
\center 
\epsfig{file=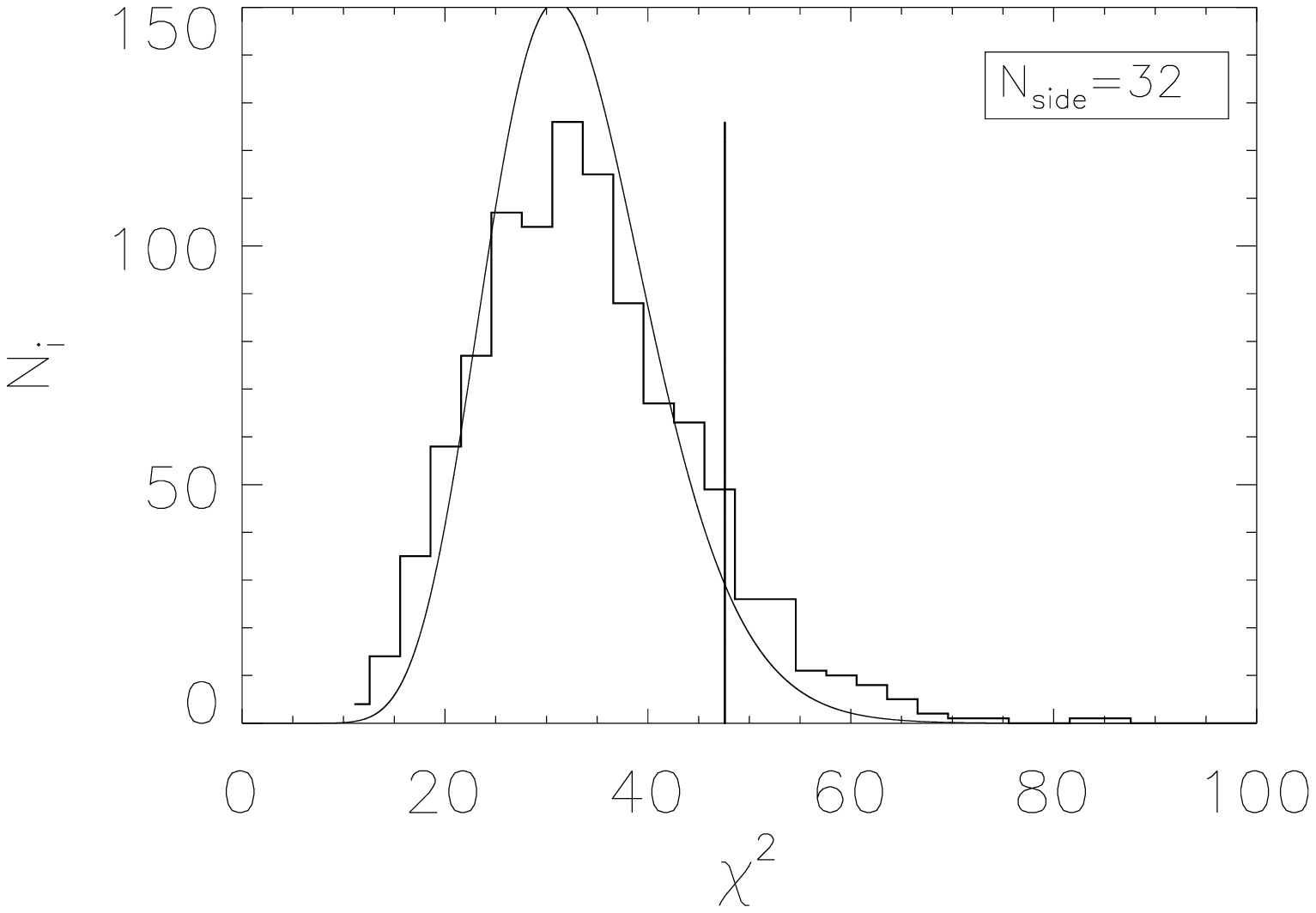,height=3.7cm,width=5.9cm}
\epsfig{file=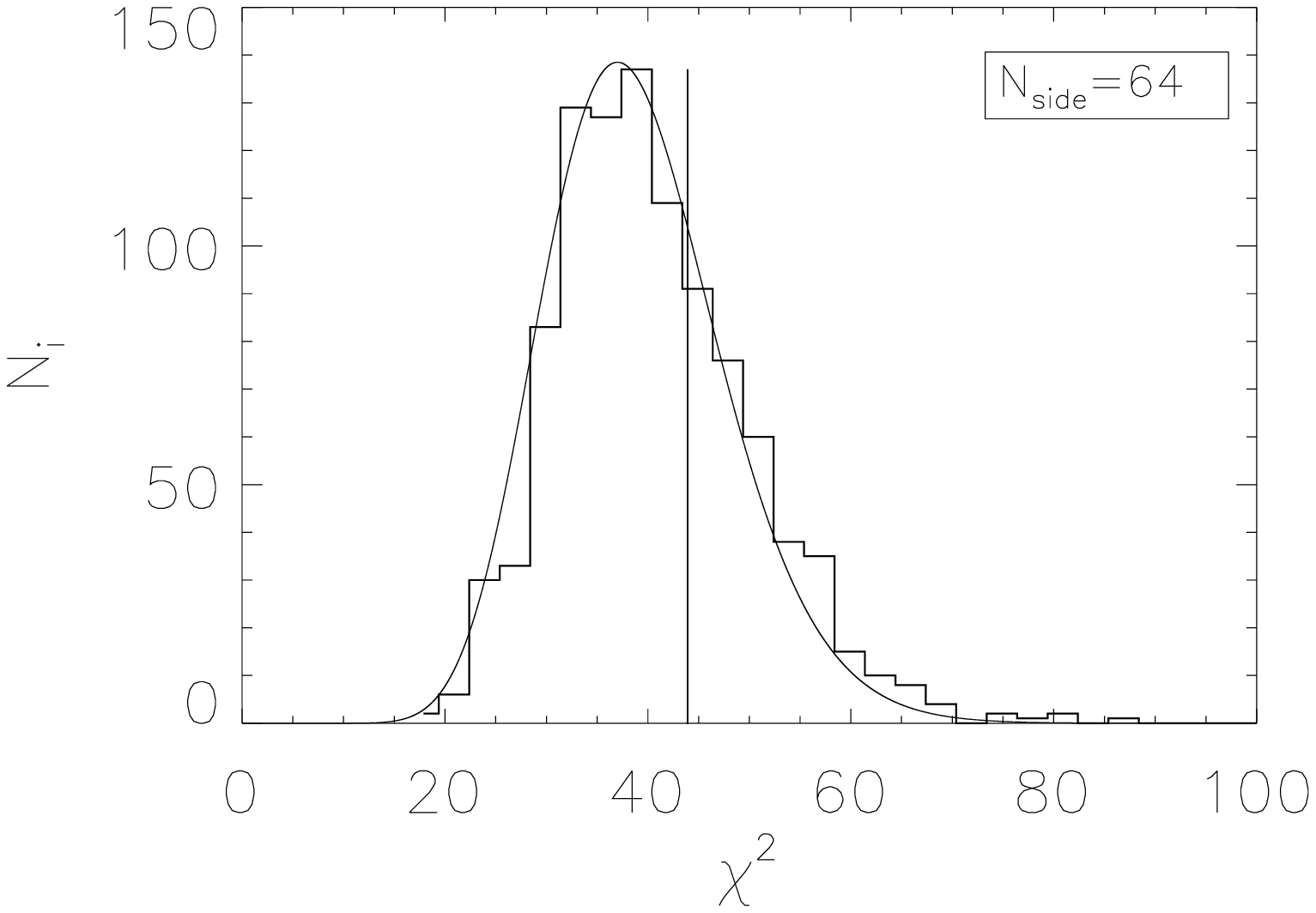,height=3.7cm,width=5.9cm}
\epsfig{file=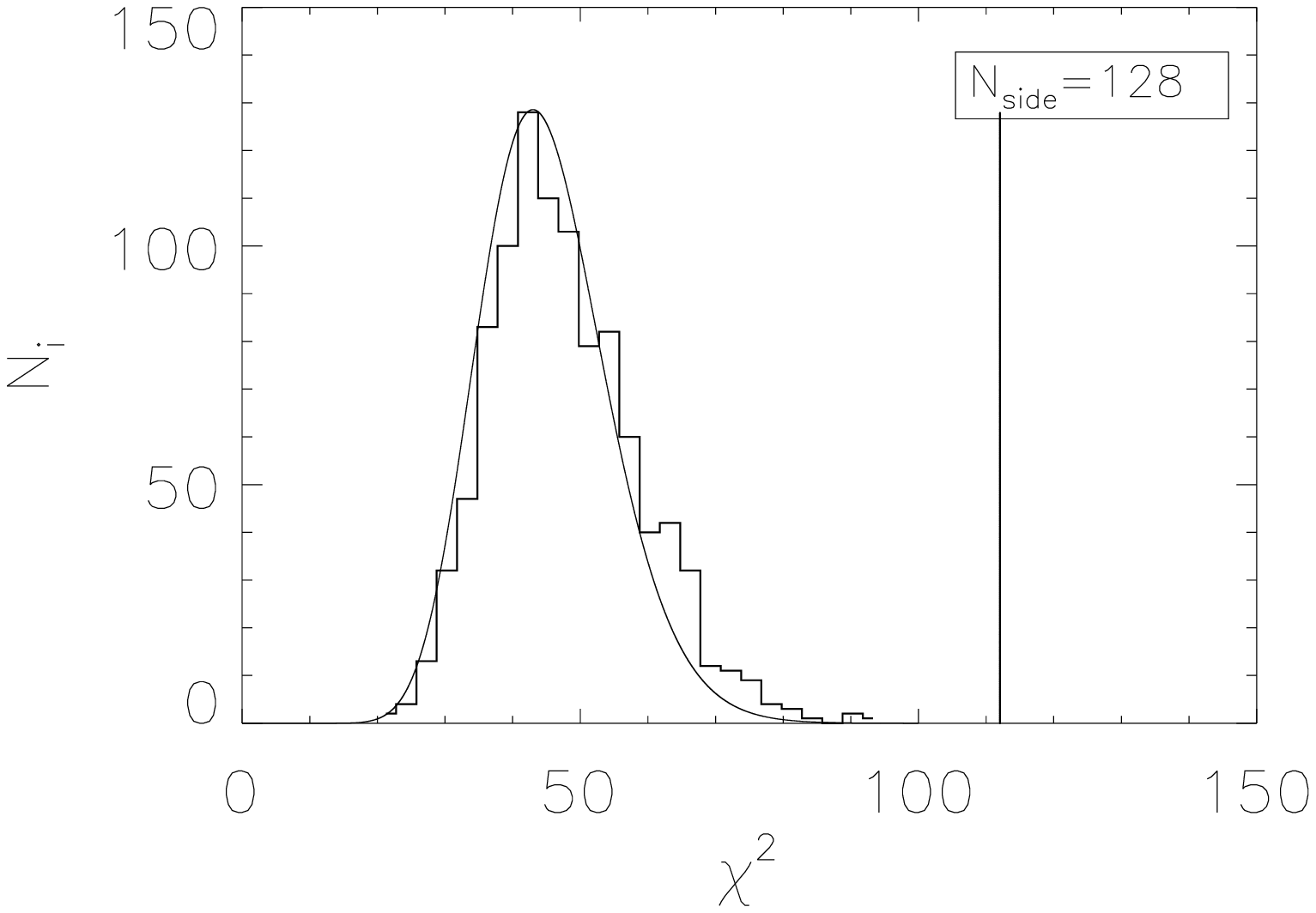,height=3.7cm,width=5.9cm}
\caption{
  Distribution of the $\chi^2$ values
  from the Minkowski Gaussianity analysis for the Archeops 143K03 noise
  map at $N_{side}=32$, $N_{side}=64$, and $N_{side}=128$. Vertical
  lines show data, the histograms corresponds to sets of
  $10^3$ noise simulations, and the solid lines are the $\chi^2$
  distribution with $3n_{th}$ degrees of freedom.}
\label{nk03gauss}
\end{figure*}
\begin{figure*}[t]
\center 
\epsfig{file=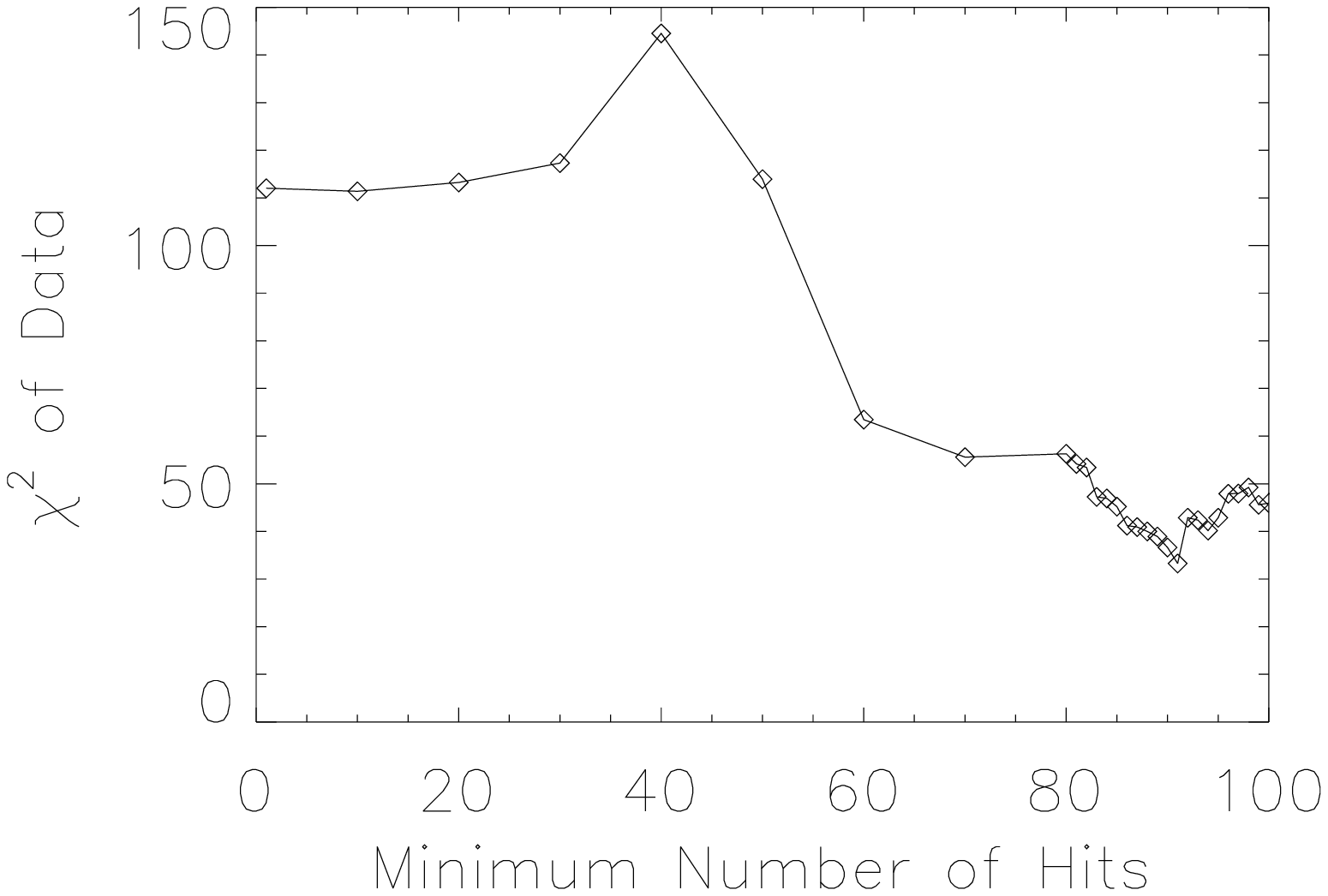,height=3.7cm,width=5.9cm}
\epsfig{file=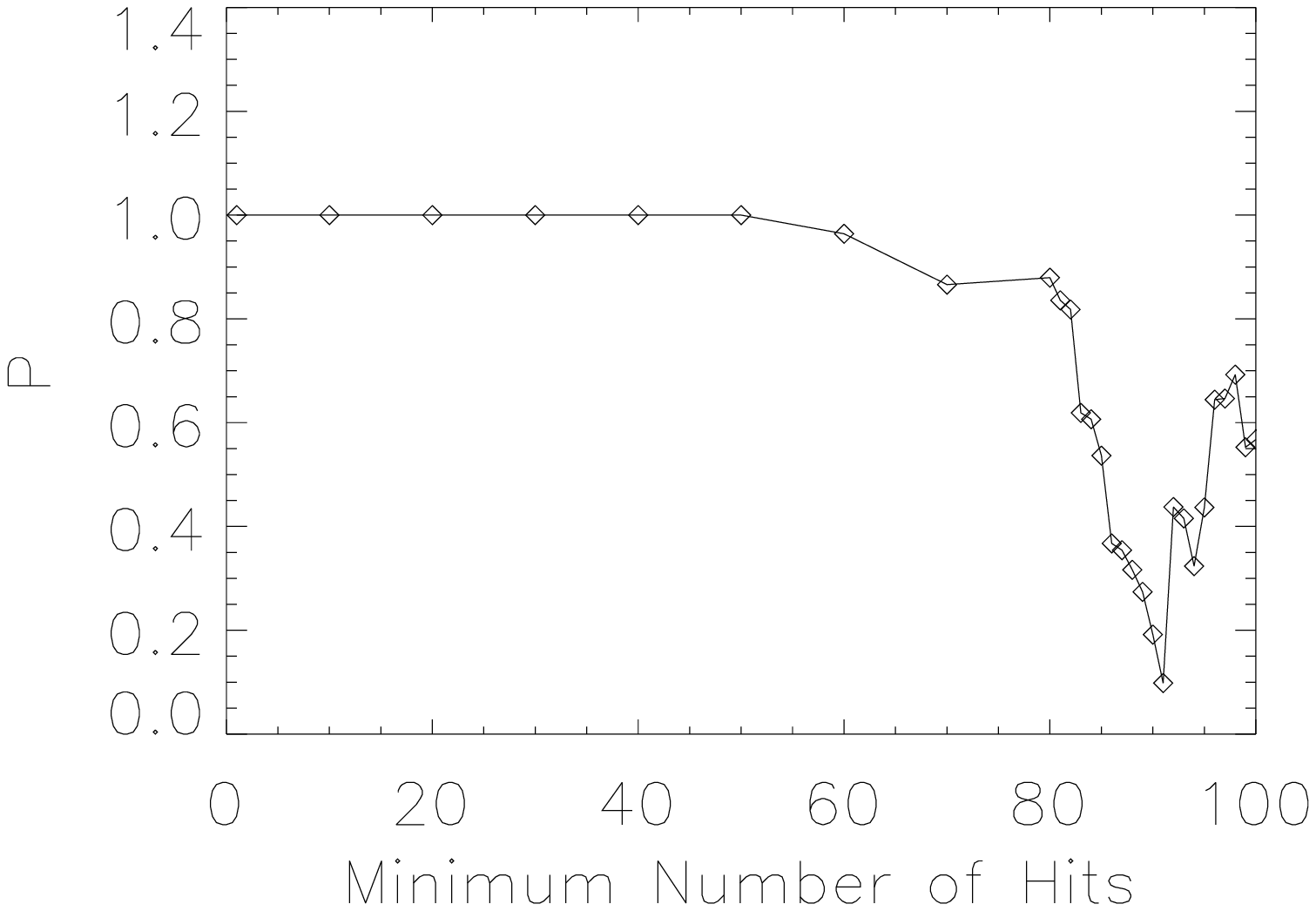,height=3.7cm,width=5.9cm}
\epsfig{file=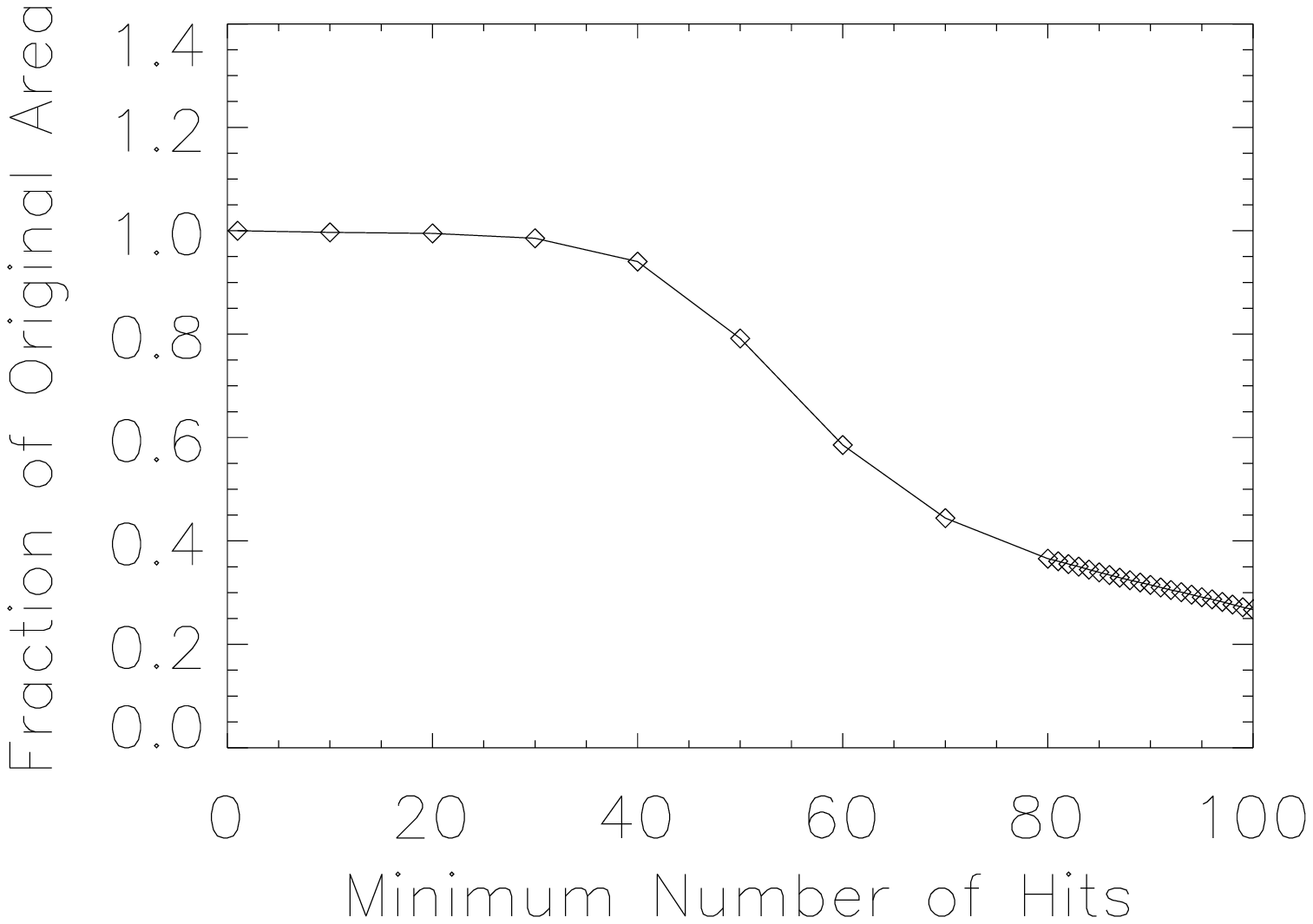,height=3.7cm,width=5.9cm}
\caption{{\it From left to right,} $\chi^2$ statistic of the 143K03
  noise map, its cumulative probability, and the fraction of available
  area for different values of the minimum number of hits at 
  $N_{side}=128$. This map becomes compatible with Gaussianity when
  the minimum number of hits is approximately 90.}
\label{chivsnhitsmin143K03noise}
\end{figure*}
From \citet{archeops_tecnical} it is expected that the noise at a
given pixel behaves as Gaussian when the pixel has been observed a
sufficiently large number of times, about 100 times. In order to check
the noise behaviour, we perform a Gaussianity analysis of the Archeops
instrumental noise. This is needed to exclude from the main
Gaussianity analysis the possible scales and regions of the sky where
the Archeops noise is non-Gaussian. 
\subsection{Two bolometer analysis}
A simple way to analyse the instrumental noise is to subtract the data
of two different bolometers at the same frequency so that the sky
signal is removed in the final map. We used the data collected by two
different Archeops bolometers at 143 GHz: 143K03 and 143K04. As the
CMB contribution is the same for both bolometers but the noise is not,
after subtraction and assuming no systematic errors only the noise
contributions remain.  Defining $d_{K03}$ as the data collected by
143K03 and $d_{K04}$ the data collected by 143K04, the map that we
analysed is
\begin{equation}
d=d_{K03}-d_{K04} \approx  n_{K03}-n_{K04}.
\label{noisediffdata}
\end{equation}
We analysed the difference data maps given by Eq. \ref{noisediffdata}
for three different resolutions: $N_{side}= 32$, $N_{side}= 64$, and
$N_{side}= 128$. To construct simulations of the difference map, we
used a set of $10^3$ Gaussian signal simulations and $10^3$ Gaussian
noise simulations for each bolometer and each resolution. These
simulations are used to compute the mean value $\langle v_i \rangle$
and the covariance matrix $C_{ij}$ of the Minkowki functionals as
described in Sect.~\ref{methodology}.
 
In Fig. \ref{d1minusd2gauss} we present, from left to right, the
Gaussianity analysis of the difference maps at the three resolution
considered, $N_{side}= 32$, $N_{side}= 64$, and $N_{side}= 128$. The
histograms correspond to the values of the $\chi^2$ obtained for
$10^3$ simulations of difference maps. The vertical lines correspond
to the value of $\chi^2$ for the Archeops data. The solid lines are
the theoretical $\chi_{3n_{th}}^2$ distribution normalised to the
number of simulations and the size of the binned cell.

From this analysis we can see that the data given by Eq.
\ref{noisediffdata} becomes non-Gaussian at high resolution.  This is
most probably due to highly noisy pixels in the difference maps
corresponding to regions of the sky observed with little redundancy.
A previous analysis \citep{archeops_tecnical} has shown that the
Archeops instrumental noise in the map domain is Gaussian distributed
for a given pixel when this pixel has been observed a significant
amount of time, typically above a few hundred independent observations
(hits) per pixel, although no precise estimate of the required number
of hits per pixel was given.

This can be easily done using the statistical tools presented in this
paper.  For this purpose we perform our analysis excluding highly
noisy pixels defined as those pixels presenting a number of hits below
a given threshold.  We computed the $\chi^2$ statistic of the data
(Eq. \ref{noisediffdata}), its cumulative probability, and the
remaining area using different thresholds for the number of hits.  By
comparing to the Gaussian simulations we observe that for both
$N_{side}= 64$ and $N_{side}= 128$, increasing the threshold of the
number of hits implies that the data becomes compatible with
Gaussianity. In particular, for $N_{side}=64$ the data become
compatible with Gaussianity when pixels with fewer than 250 hits are
removed (leaving $46\%$ of the original area) and for $N_{side}=128$
the data start to become compatible with Gaussianity when pixels with
fewer than 150 hits are removed (leaving $13\%$ of the original area).
%
%
\subsection{Single bolometer analysis}
From the above analysis we have proved that highly noisy pixels are
responsible for the non-Gaussianity of the Archeops noise at high
resolution. In order to obtain proper limits for the $f_{nl}$
parameter we need to exclude these pixels in our final analysis of the
$143K03$ map.

To avoid any contamination from the $143K04$ noise map when characterizing
the noise map for the 143K03 bolometer the latter was computed using the WMAP
data as follows
\begin{equation}
n_{K03} = d_{K03} - WMAP_{K03}*f
\label{noiseK03}
\end{equation}
where $WMAP_{K03}$ is the combined WMAP data computed as described in
\cite{curto_32} and converted into an Archeops map as described in
Sect.~\ref{model}. $f$ is an intercalibration parameter between WMAP
and Archeops, $f = 1/1.07$, as given by \citet{tristram_cl}.  In this
case, as the noise on the WMAP data at the considered map resolutions
is negligible compared to the Archeops noise, we expect no
contamination.

We analysed the noise map given by Eq. \ref{noiseK03} for three
different pixel resolutions: $N_{side}= 32$, $N_{side}= 64$, and
$N_{side}= 128$. For each resolution, we also analysed a set of $10^3$
Archeops noise simulations $n_{K03}$. The results are presented in
Fig.  \ref{nk03gauss}. The histograms correspond to the $\chi^2$ of
the Minkowski functionals of a set of $10^3$ $n_{K03}$ Gaussian
simulations, vertical lines correspond to the data maps given by
Eq. \ref{noiseK03} and the solid lines are the $\chi^2_{3n_{th}}$
distributions.  We can see that for low resolution ($N_{side}= 32$ and
$N_{side}= 64$) the noise map is compatible with the Gaussian noise
simulations. Therefore for these two resolutions we can use the full
available area in the $f_{nl}$ analysis.
%

In the case of $N_{side}= 128$ the noise map is not compatible with
Gaussianity. As above, we reanalysed this noise map removing highly
noisy pixels with a number of hits below a given threshold. This
analsysis is presented in Fig.~\ref{chivsnhitsmin143K03noise} where
from left to right we plot the $\chi^2$ statistic of the 143K03 noise
map, its cumulative probability, and the fraction of available area
for different values of the minimum number of hits.  We can see that
the Archeops noise for the $143K03$ bolometer becomes clearly Gaussian
when we remove pixels with a number of hits lower than 90. This means
that the area where the noise is Gaussian is $32\%$ of the original
area at $N_{side}=128$.
In conclusion, the regions of the sky that we can use for the
Gaussianity analysis are the whole initial area at $N_{side}=32$ and
$N_{side}=64$ and $32\%$ of the initial area at $N_{side}=128$.

The $143K04$ bolometer has a qualitatively similar behaviour for the
noise analysis. That is, the noise map is compatible with Gaussianity
at low resolution and is non-Gaussian at high resolution. Nevertheless
as the noise for this bolometer has more systematic errors than the
$143K03$ bolometer, the thresholds for the number of hits where the
maps become compatible with Gaussianity analysis are higher. This
implies that the available area where the noise is Gaussian is
smaller.
%
%
%
\section{Results}
\label{results}
In this section, we perform the Gaussianity analysis of the $143K03$
Archeops bolometer data in order to constrain the non-linear coupling
parameter, $f_{nl}$.  For this, we first consider realistic
non-Gaussian simulations as described in Sect.~\ref{model}. Then the
results are compared to the ones obtained for non-Gaussian simulations
for which only the Sachs-Wolfe contribution is included, as
in~\cite{curto_32}.
\subsection{Gaussianity analysis}
\begin{figure*}[t]
\center 
\epsfig{file=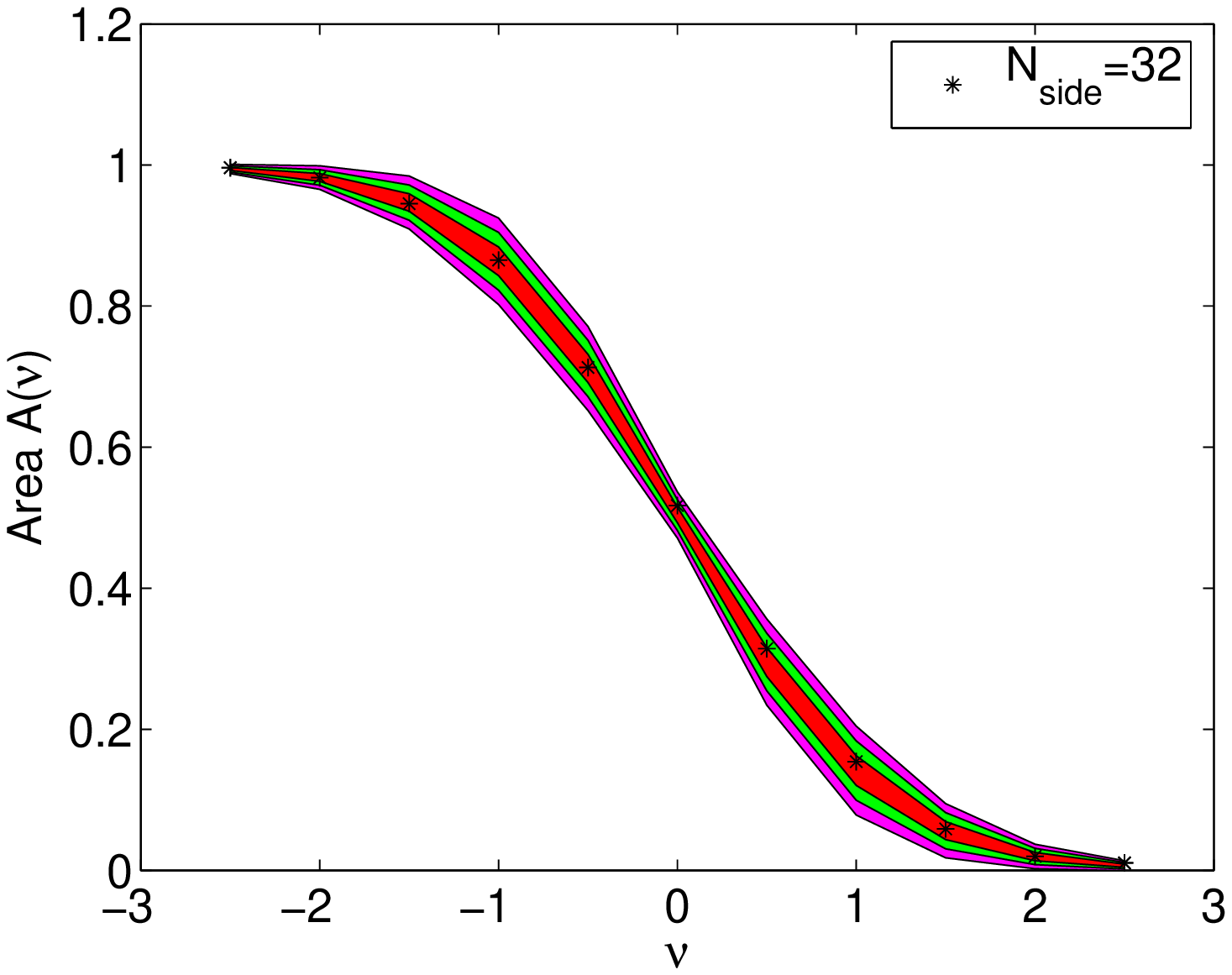,height=3.5cm,width=5.83cm}
\epsfig{file=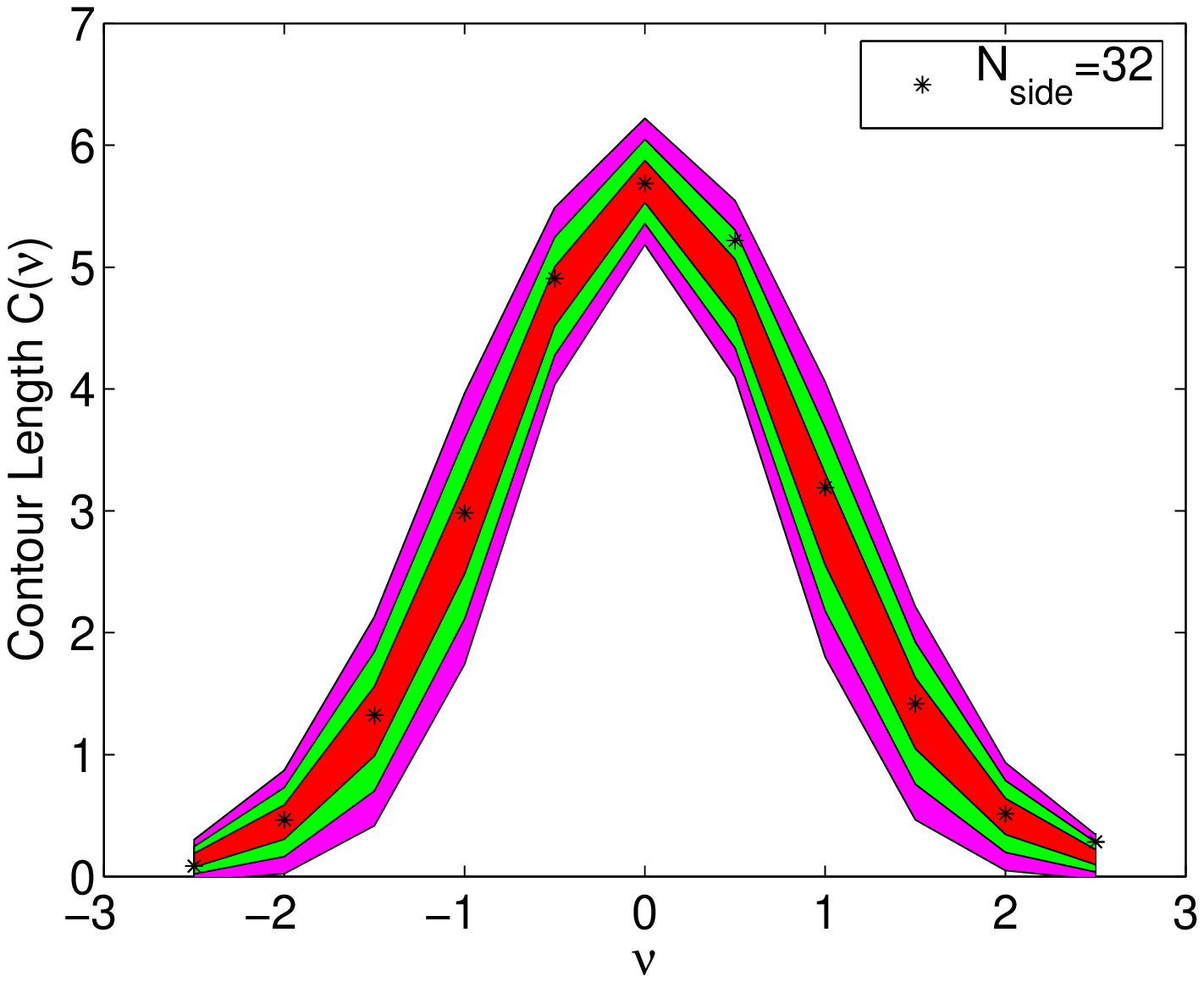,height=3.5cm,width=5.83cm}
\epsfig{file=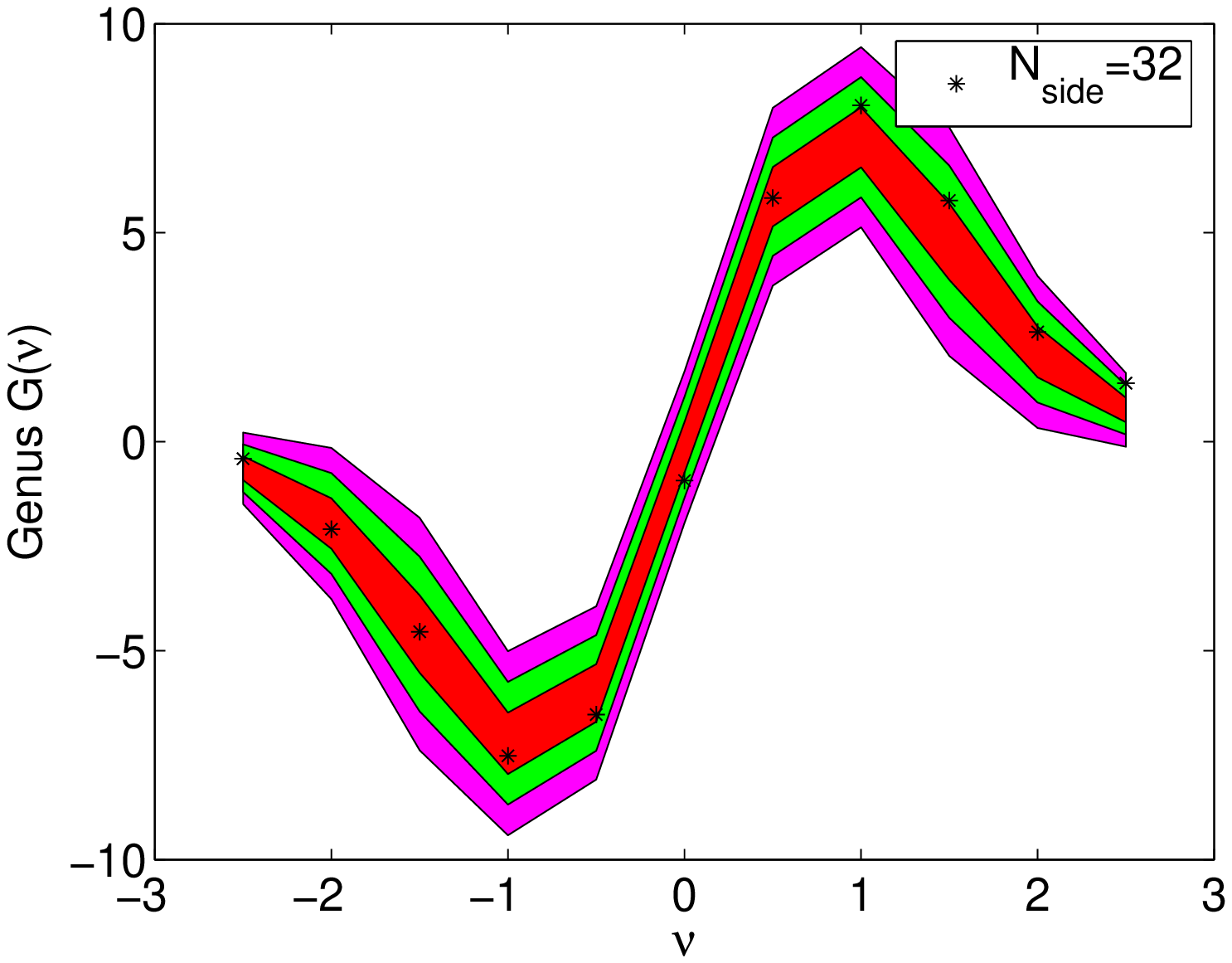,height=3.5cm,width=5.83cm}
\epsfig{file=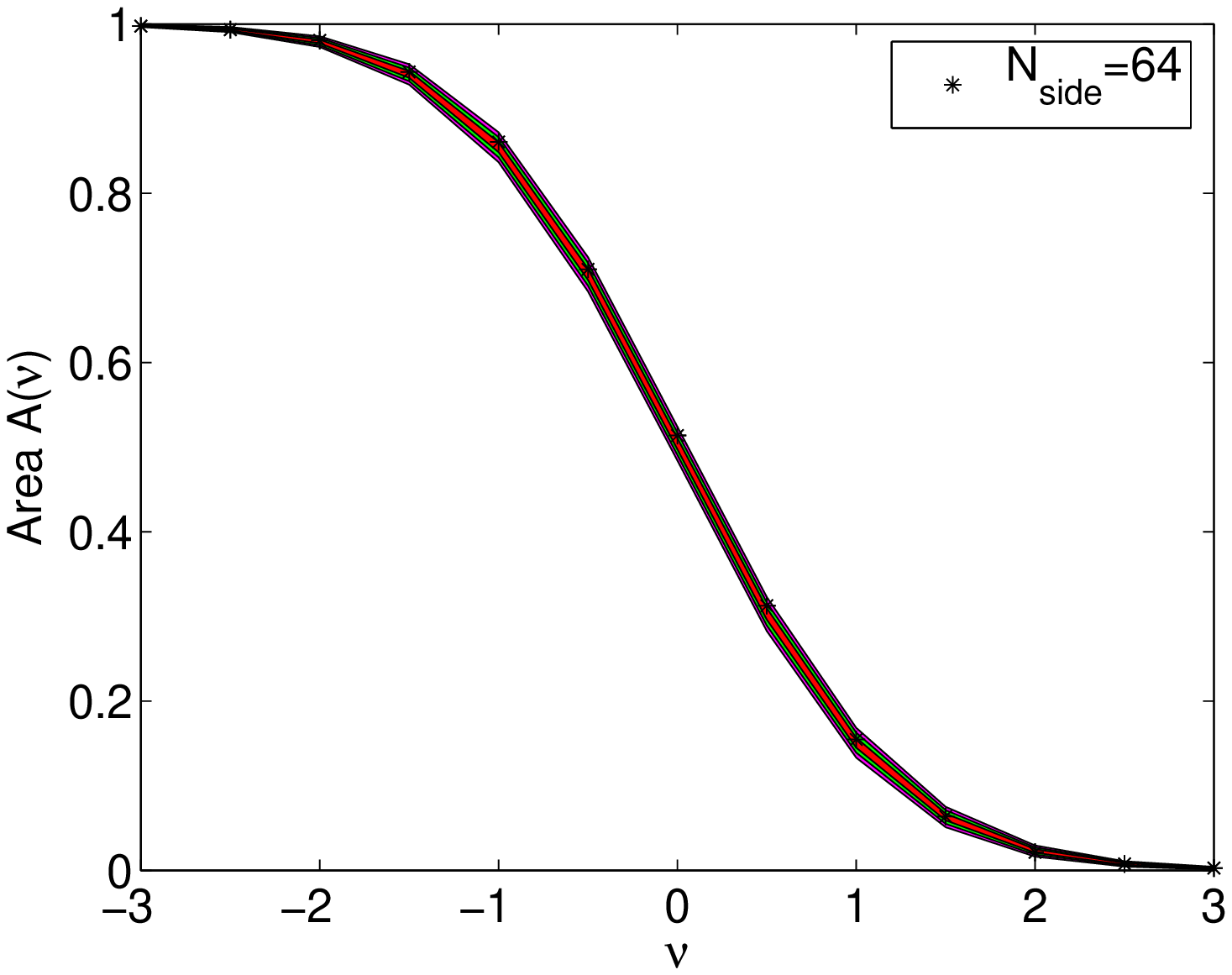,height=3.5cm,width=5.83cm}
\epsfig{file=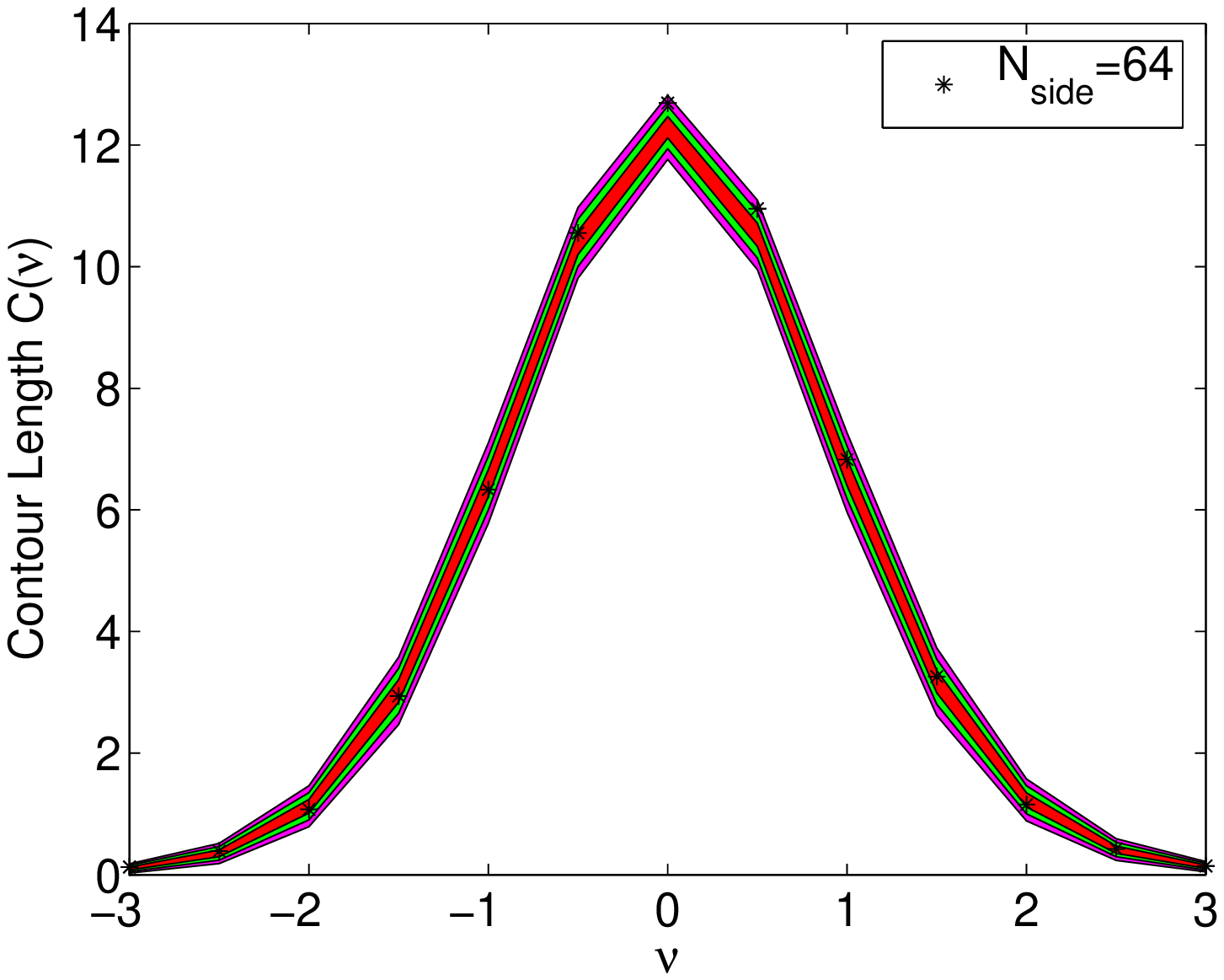,height=3.5cm,width=5.83cm}
\epsfig{file=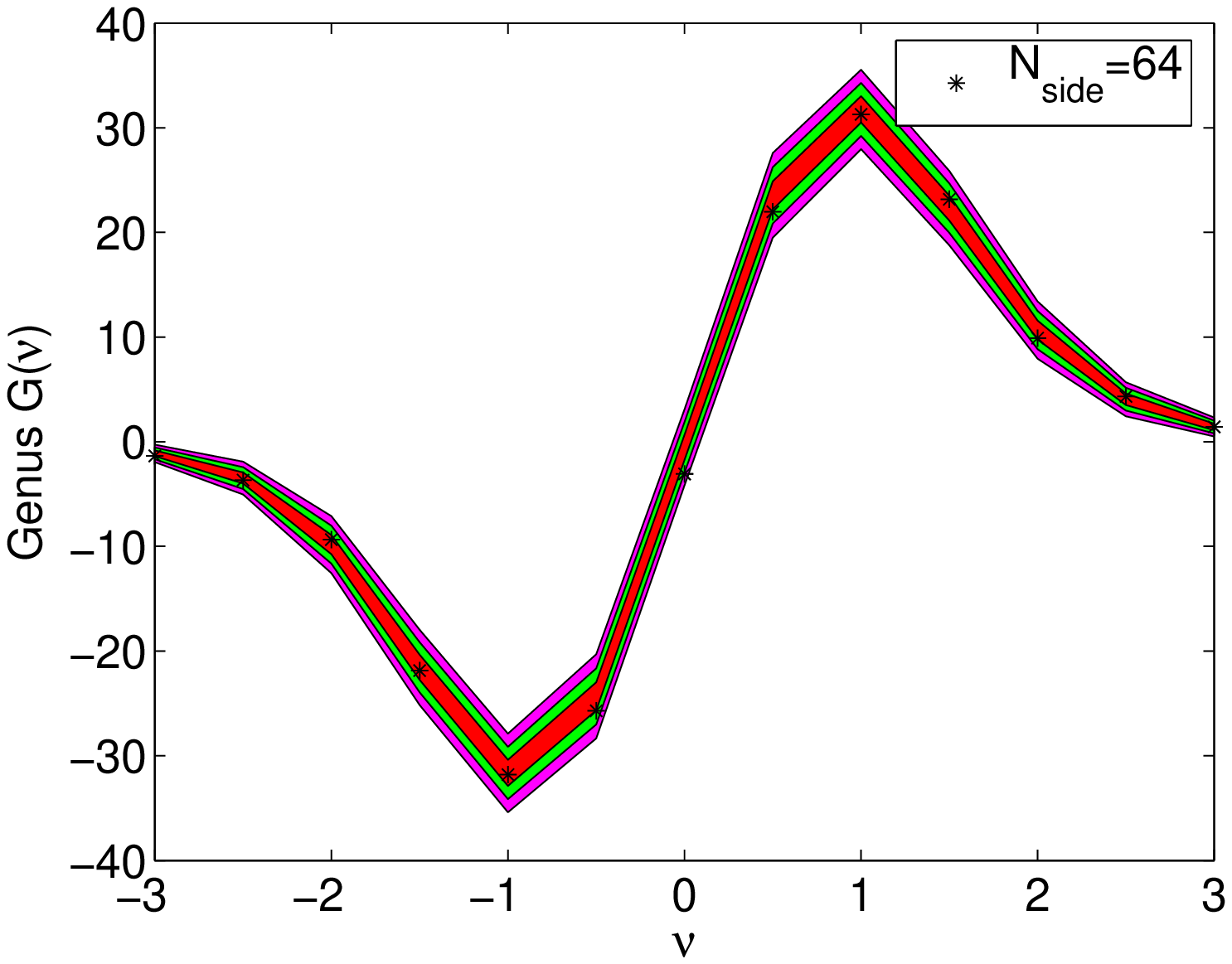,height=3.5cm,width=5.83cm}
\epsfig{file=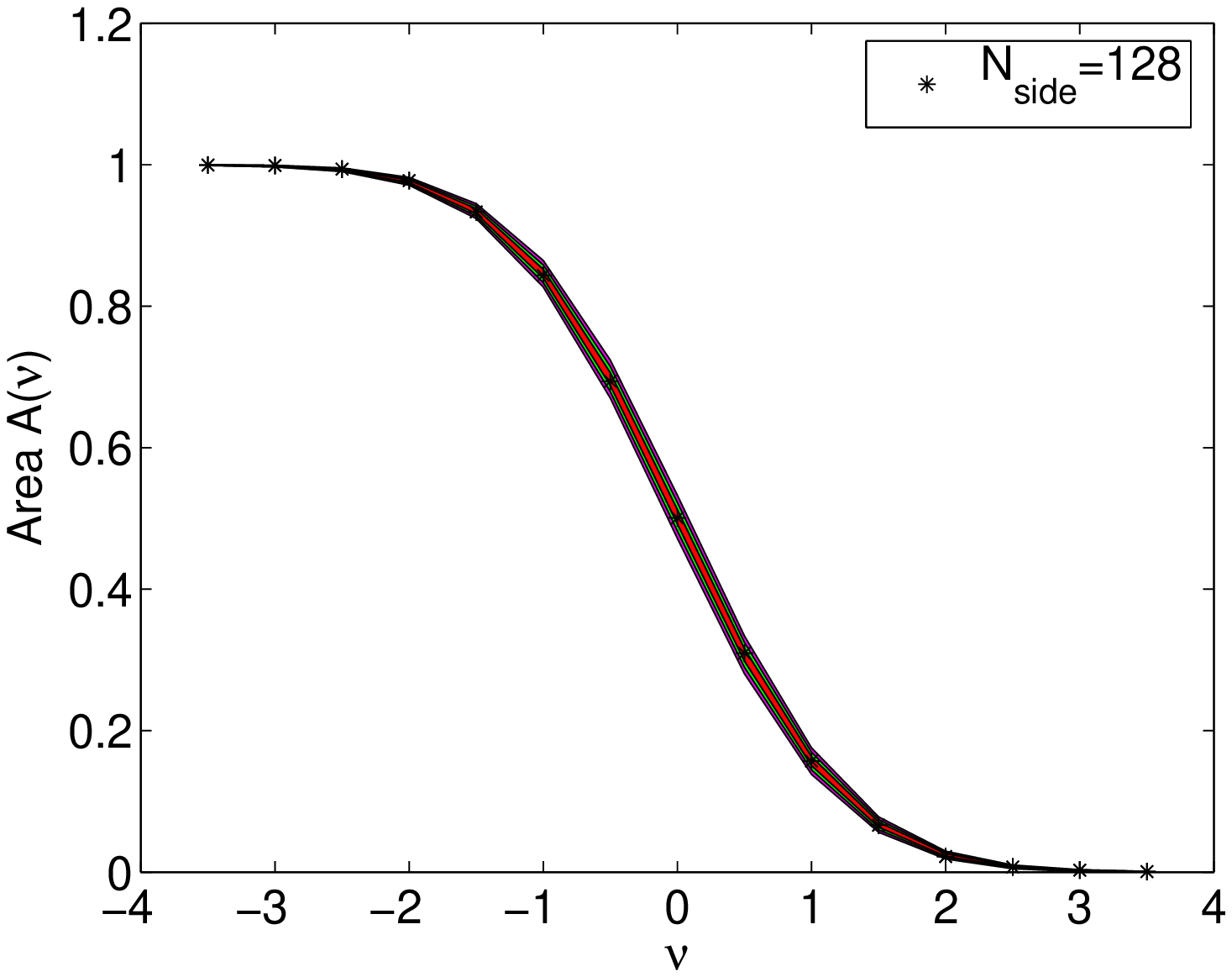,height=3.5cm,width=5.83cm}
\epsfig{file=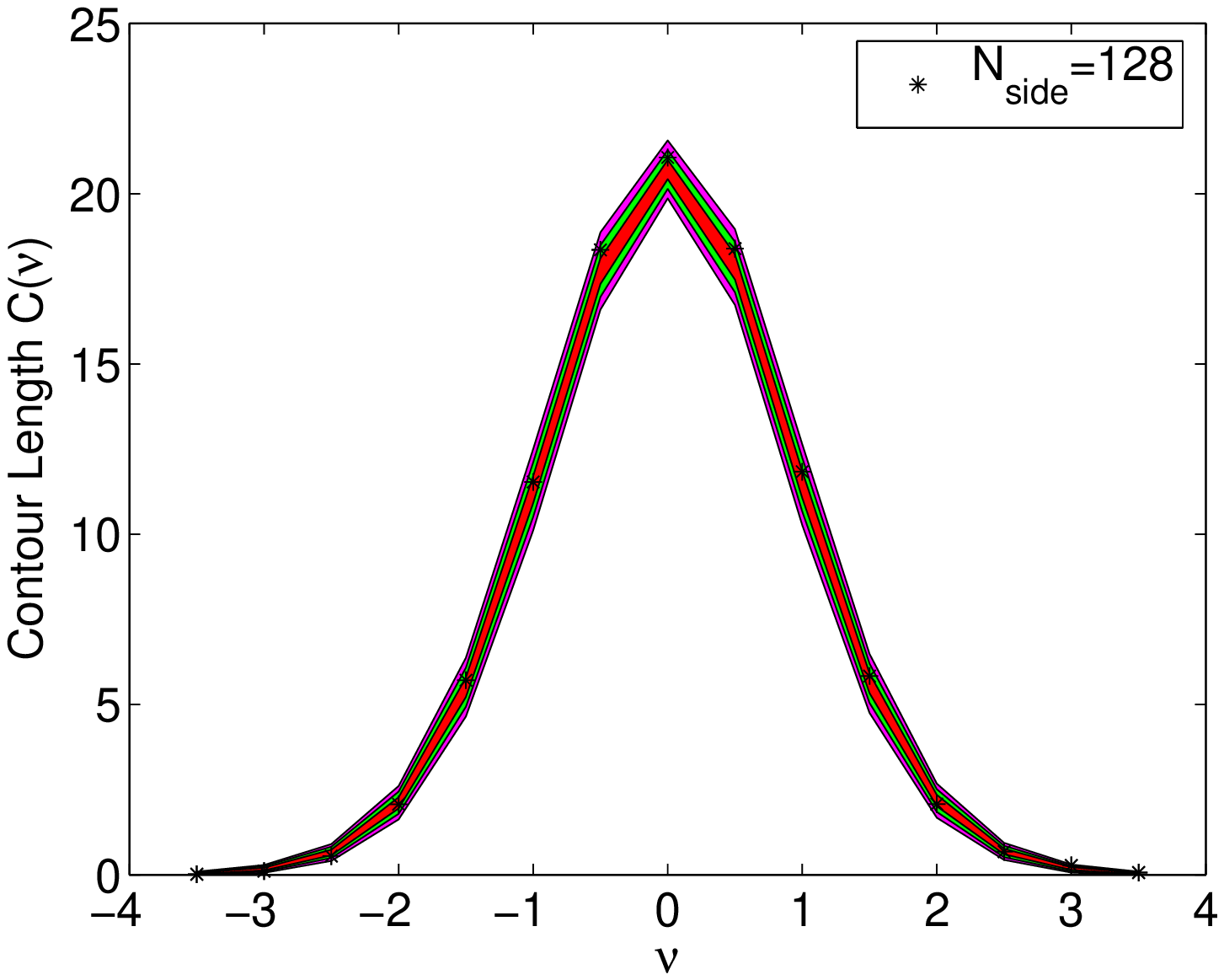,height=3.5cm,width=5.83cm}
\epsfig{file=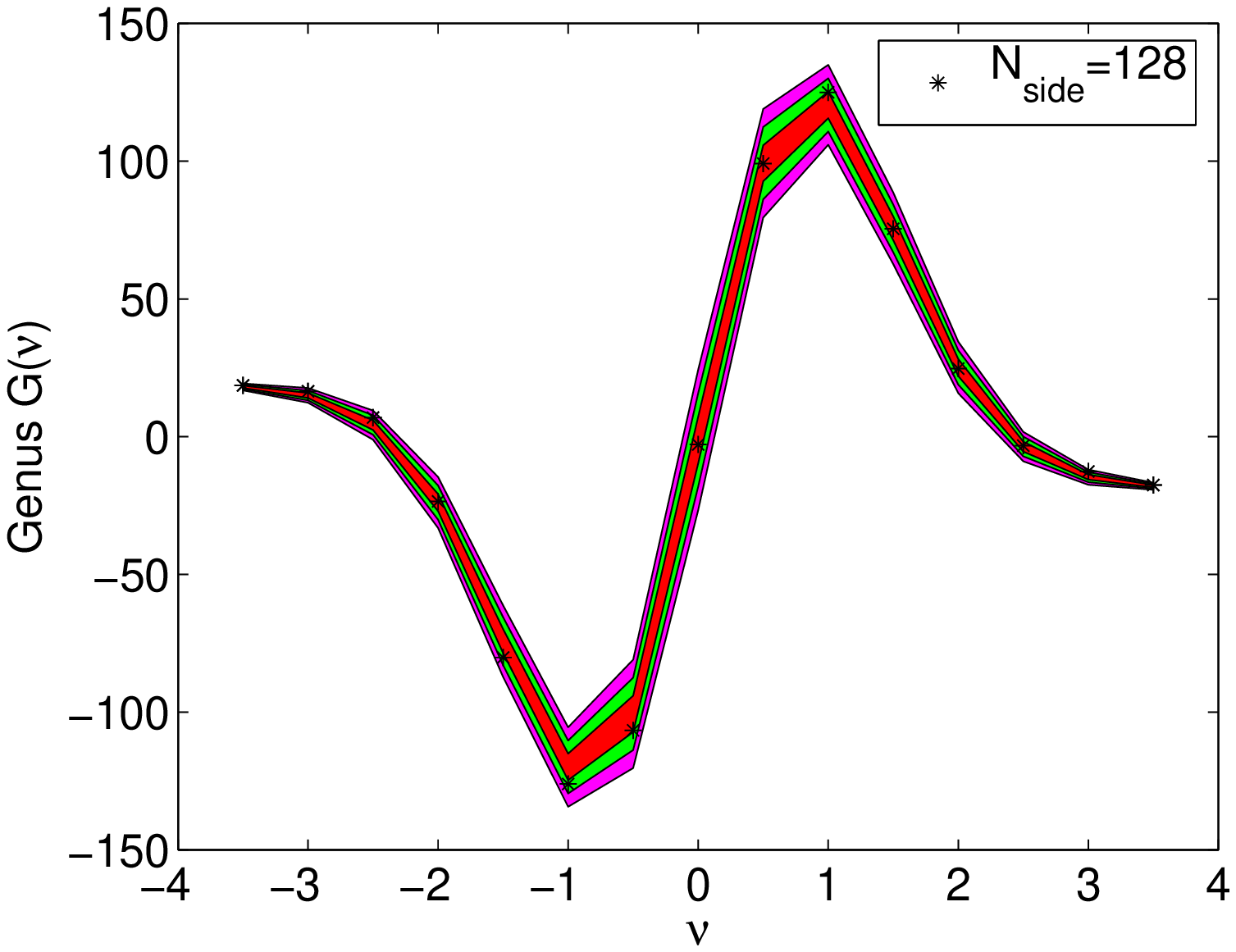,height=3.5cm,width=5.83cm}
\caption{{\it From left to right,} and {\it from top to bottom,} the
  area, contour length, and genus of the data (asterisk *) as a function
  of threshold for the 143K03 maps with $N_{side}=32$, $N_{side}=64$, and
  $N_{side}=128$. We also plot the acceptance intervals for the 68\%
  (inner in red), the 95\% (middle in green), and 99\% (outer in
  magenta) significance levels given by 10000 Gaussian simulations 
  of signal and noise.}
\label{fun143K03_ns32_ns64_ns128min90}
\end{figure*}
%
%
\begin{figure*}[t]
\center 
\epsfig{file=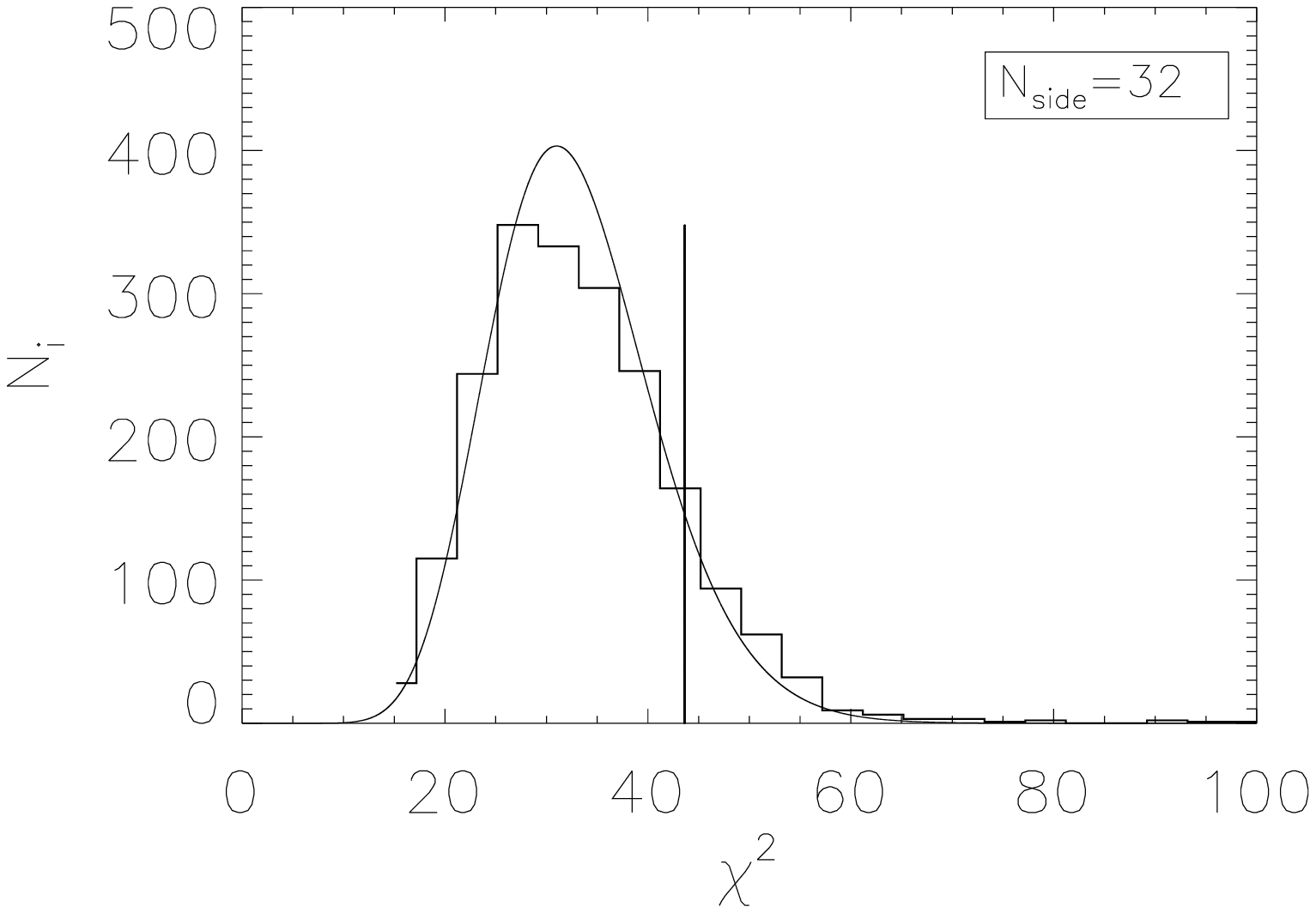,height=3.7cm,width=5.9cm}
\epsfig{file=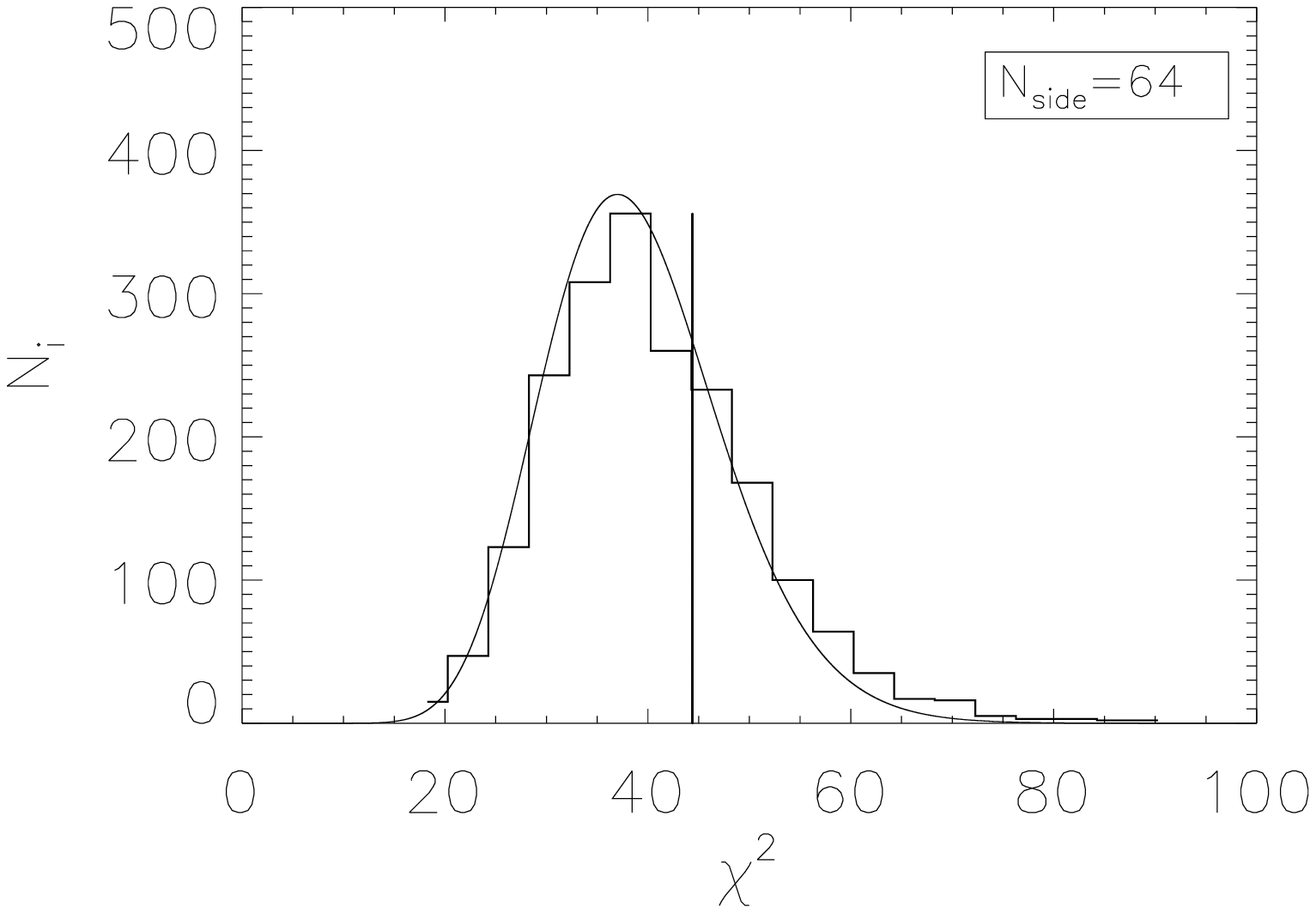,height=3.7cm,width=5.9cm}
\epsfig{file=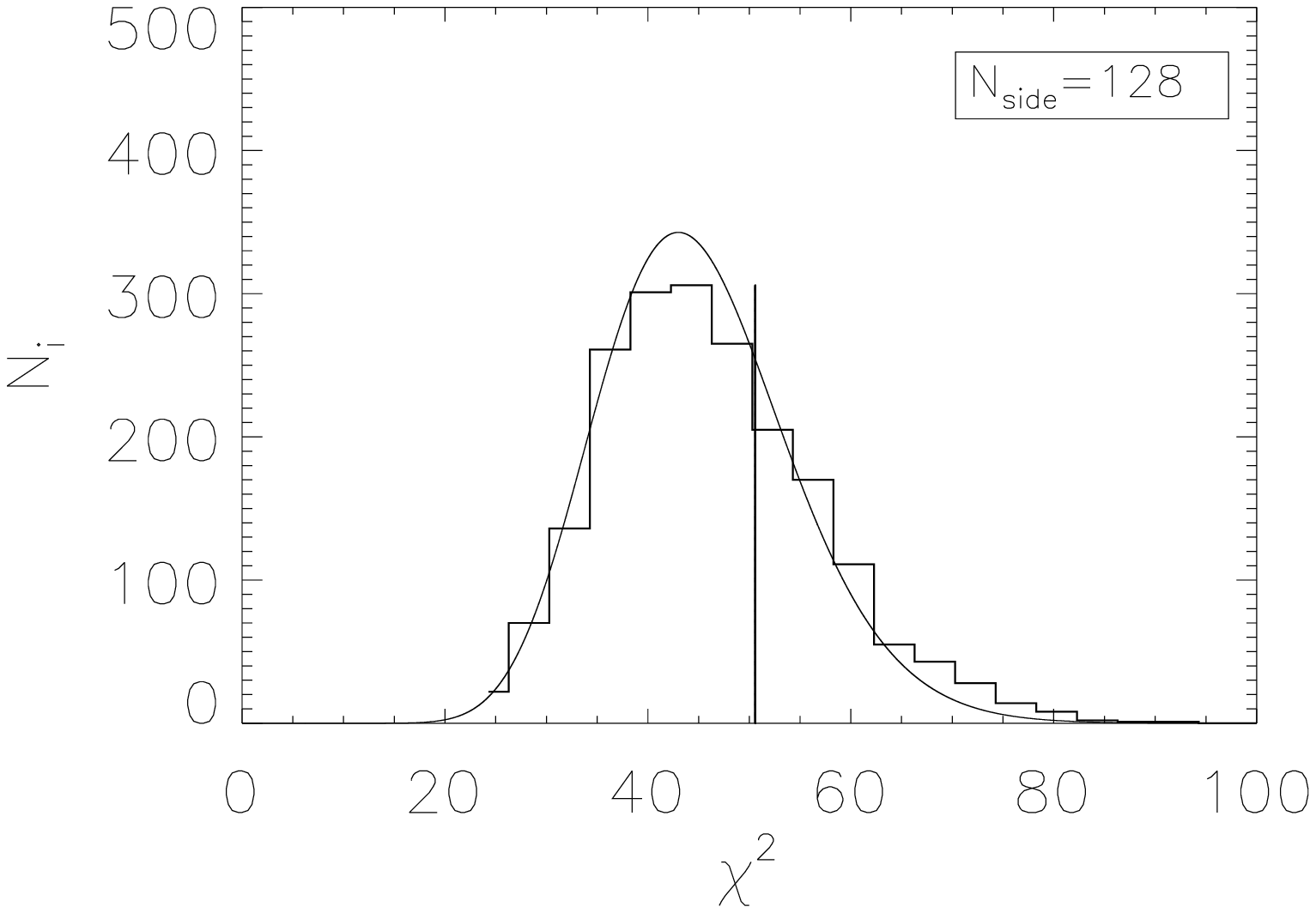,height=3.7cm,width=5.9cm}
\epsfig{file=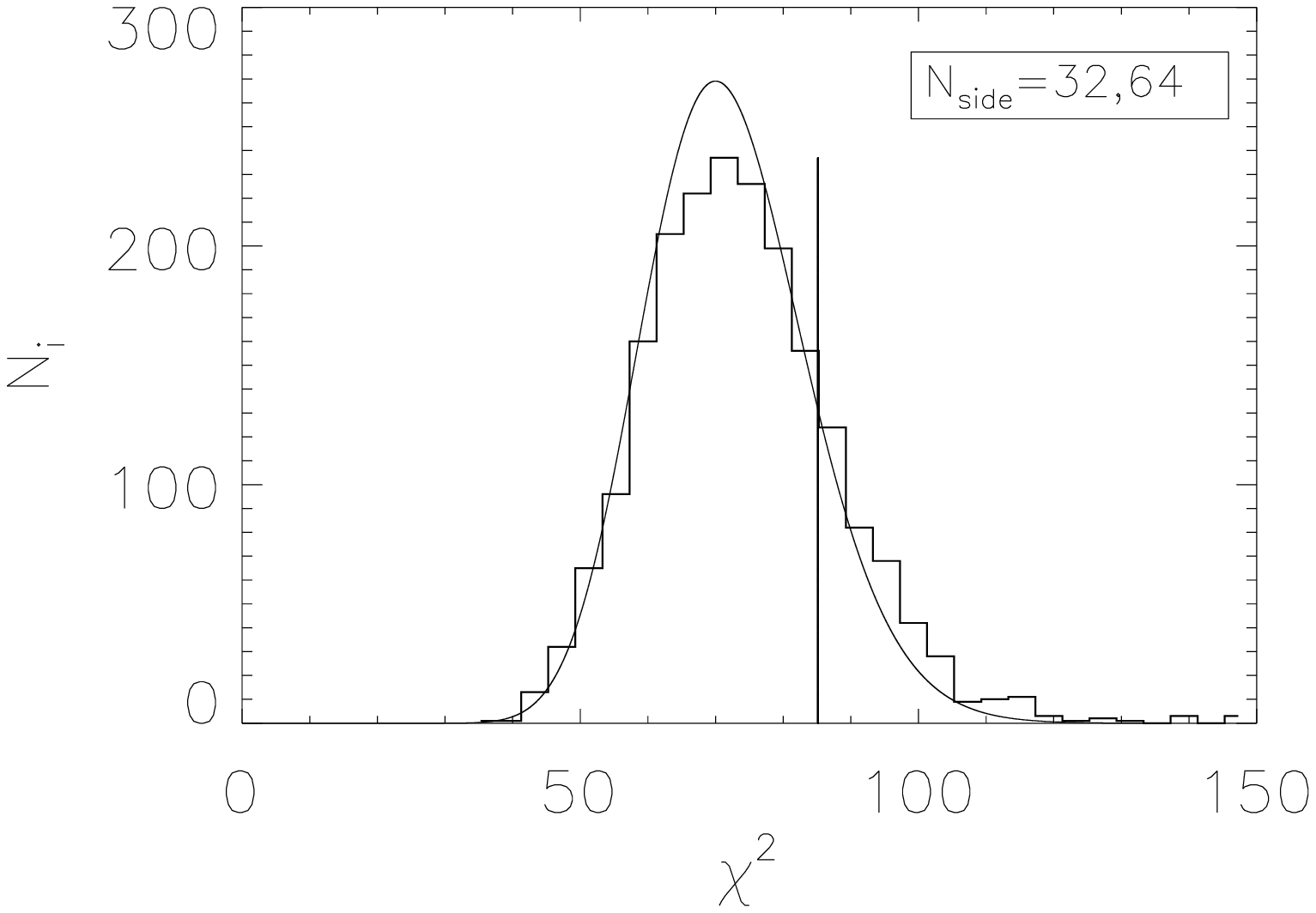,height=3.7cm,width=5.9cm}
\epsfig{file=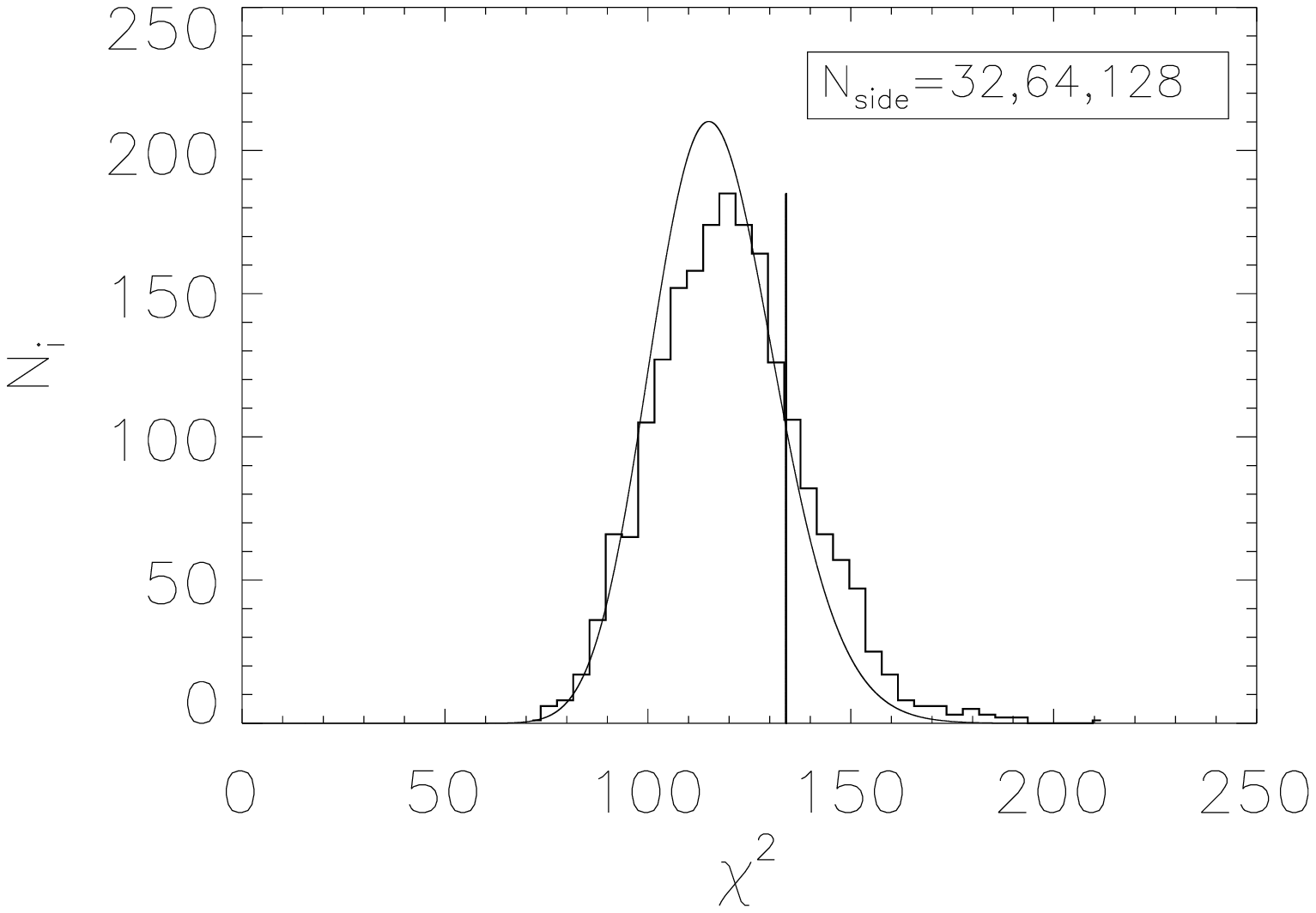,height=3.7cm,width=5.9cm}
\caption{{\it From left to right,} and {\it from top to bottom,} the
Gaussianity analysis of Archeops data maps at resolutions:
$N_{side}=32$, $N_{side}=64$, $N_{side}=128$, the combinations
$N_{side}=32,64$, and the combinations $N_{side}=32,64,128$. The
histograms correspond to the $\chi^2$ of 2000 Gaussian simulations,
the vertical lines are the $\chi^2$ of the data and the solid lines
are the expected $\chi^2$ distribution with $3N_{th}$ degrees of
freedom.}
\label{chi143K03_ns32_ns64_ns128min90}
\end{figure*}
\begin{table*}[t]
  \center
  \caption{$\chi^2$ for the three Minkowski functionals for different
    resolutions and thresholds.  \label{table_stat}}
  \begin{tabular}{ccccccc}
    \hline 
    \hline
    Resolution & Area (\%) & $\chi_{data}^2$ & Degrees Of Freedom & $\langle \chi^2 \rangle$ & $\sqrt{\langle (\chi^2 -\langle \chi^2 \rangle)^2\rangle}$ & $P(\chi^2\le \chi_{data}^2)$\\
    \hline
    32 & 100 & 43.62 & 33 & 33.55 & 10.05 & 0.868 \\
    64 & 100 & 44.39 & 39 & 40.55 & 10.33 & 0.679 \\
    128 & 32 & 50.59 & 45 & 46.29 & 10.79 & 0.688 \\
    32,64 & 100,100 & 85.14 & 72 & 73.88 & 14.29 & 0.804 \\
    32,64,128 & 100,100,32 & 134.05 & 117 & 120.30 & 18.40 &  0.791 \\
    \hline
    \hline
  \end{tabular}
\end{table*}
\begin{table*}[t]
  \center
  \caption{Best fit $f_{nl}$ of Archeops data at different
    resolutions, mean, dispersion and some percentiles of 
    the $f_{nl}$ distributions obtained from 2000 Gaussian simulations.
    \label{table_minfnl}}
  \begin{tabular}{ccc|cc|cc|cc}
    \hline 
    \hline
    Resolution & Area (\%) & Best Fit $f_{nl}$  & $\langle f_{nl} \rangle$& $\sqrt{\langle f_{nl}^2 \rangle-\langle f_{nl} \rangle^2}$ & $X_{0.160}$& $X_{0.840}$ & $X_{0.025}$& $X_{0.975}$ \\ 
    \hline
    32 & 100 & 90 & 48 & 702 & -670 & 785 & -1370 & 1445 \\
    64 & 100 & 75 & -29 & 645 & -665 & 615 & -1315 & 1270 \\
    128 & 32 & -45 & 112 & 880 & -810 & 1000 & -1665 & 1805 \\
    32,64 & 100,100 & 70 & -3 &  550 & -525 & 535 & -1130 & 1070 \\
    32,64,128 & 100,100,32 & 70 & 19 & 503 & -470 & 520 & -990 & 1005 \\
    \hline
    \hline
  \end{tabular}
\end{table*}
\begin{figure*}[t]
\center 
\epsfig{file=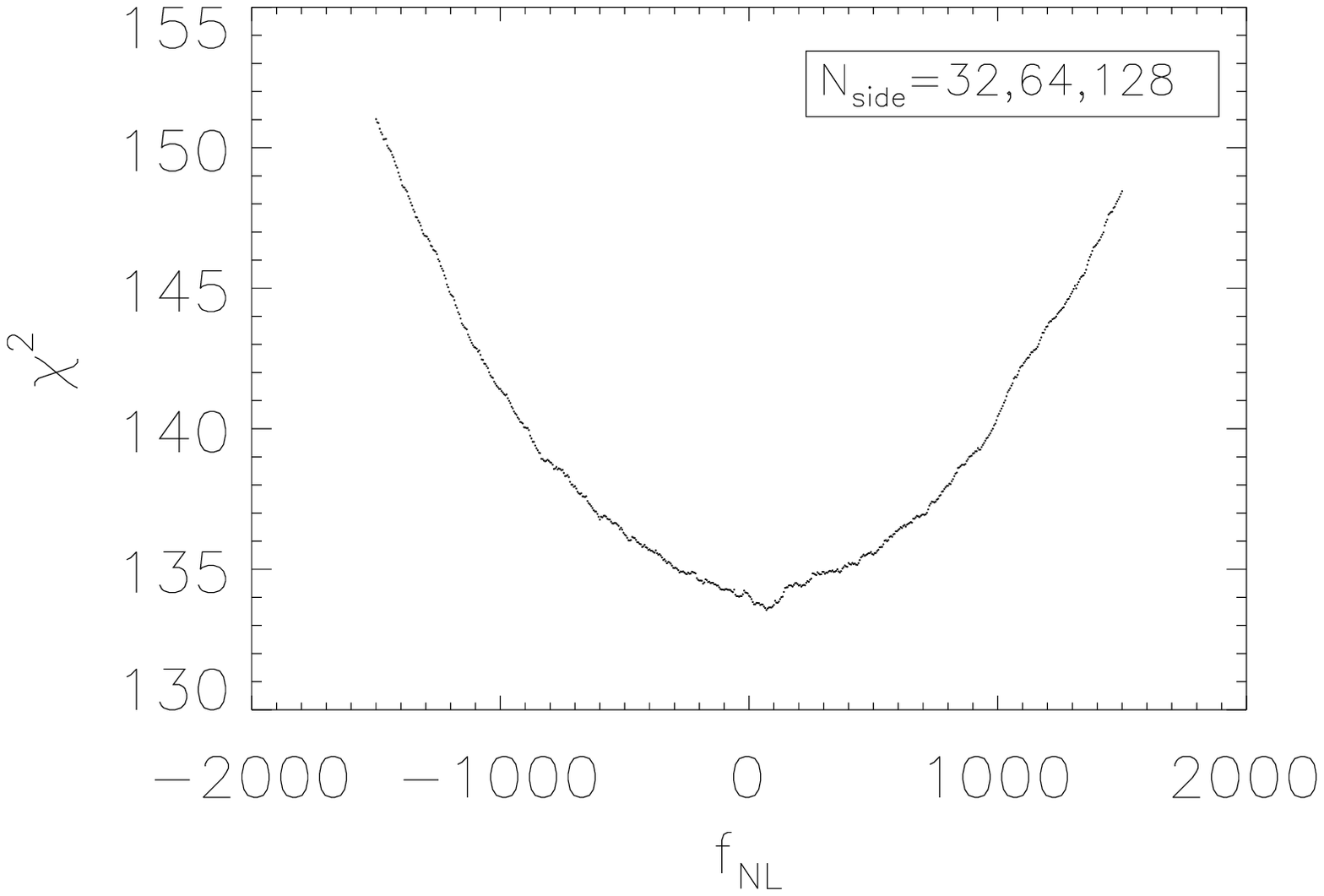,height=4.5cm,width=8cm}
\epsfig{file=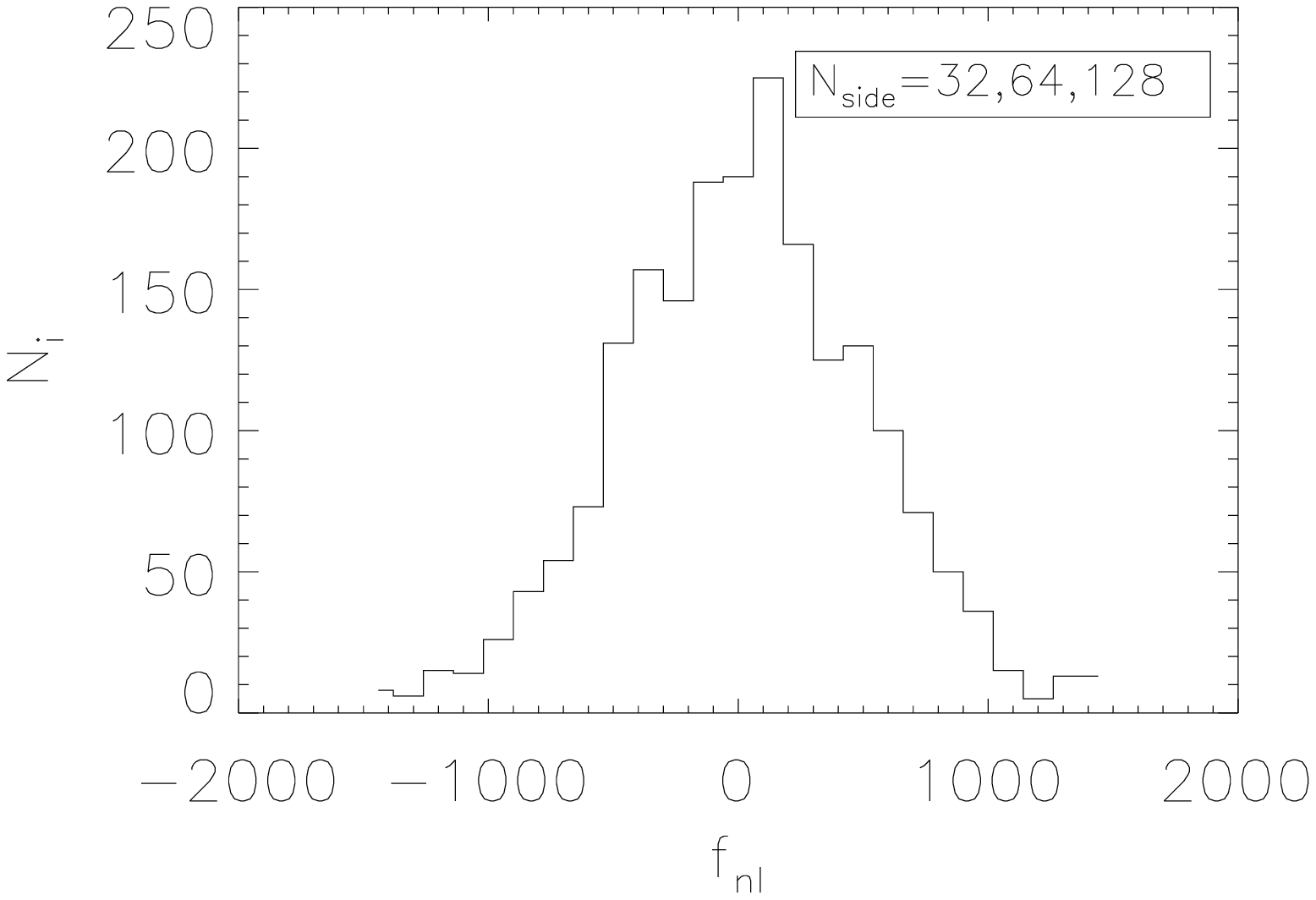,height=4.5cm,width=8cm}
\caption{{\it From left to right,} $\chi^2(f_{nl})$ of Archeops data
  vs $f_{nl}$ and the best fit value of $f_{nl}$ for a set of 2000
  Archeops Gaussian simulations ($s_{k03}+n_{k03}$) combining the maps
  at resolutions
  $N_{side}=32,64,128$.\label{chi143K03fnl_ns32_ns64_ns128min90}}
\end{figure*}
For this analysis we apply the statistical methods described in
Sect.~\ref{methodology} to the Archeops 143K03 bolometer data.  Figure
\ref{fun143K03_ns32_ns64_ns128min90} shows the Minkowski functionals
as a function of threshold for maps with $N_{side}=32$, $N_{side}=64$,
and $N_{side}=128$. In addition, we plot the acceptance intervals for
the 68\% (inner), the 95\% (middle), and 99\% (outer) significance
levels obtained from 10000 noise and signal Gaussian simulations of
the Archeops 143K03 data as described in Sect.~\ref{datamethod}. For
all the analysed cases the data are compatible with Gaussianity at
least at the 99\% significance level.

In Fig. \ref{chi143K03_ns32_ns64_ns128min90} we present the results of
the Gaussianity analysis of the Archeops 143K03 bolometer data at
different resolutions using the $\chi^2$ test described by
Eqs. \ref{chi2} and \ref{bigchi2}. In particular, we analysed the data
at $N_{side}=32$ (11 thresholds, 3 functionals), $N_{side}=64$ (13
thresholds, 3 functionals), $N_{side}=128$ (15 thresholds, 3
functionals), the combination of the data at $N_{side}=32$ and
$N_{side}=64$ (total of 24 thresholds, 3 functionals), and the
combination of the data at $N_{side}=32$, $N_{side}=64$, and
$N_{side}=128$ (total of 39 thresholds, 3 functionals). To avoid
confusion by the non-Gaussianity of the Archeops noise at high
resolutions, the high noise pixels were excluded as discussed in the
previous Section. The histograms in the plot correspond to the
analysis of 2000 Archeops signal plus noise Gaussian simulations at
each resolution, the solid lines correspond to the expected
$\chi_{3N_{th}}^2$ distribution, and the vertical lines correspond to
$\chi^2$ values of the data at each resolution.  For all the
resolutions and combination of resolutions, the Archeops data are
consistent with Gaussianity as expected from the previous figure.  In
Table \ref{table_stat} we present the $\chi^2$ value of the data for
the different cases that we have analysed, the total number of degrees
of freedom, the mean and the dispersion of the $\chi^2$ value of the
Gaussian simulations, and the cumulative probability of the data
computed from the distribution given by the simulations.

If we use all the available area at $N_{side}=128$ we find that
Archeops data are not compatible with Gaussianity. In particular, we
have obtained $P(\chi^2\le \chi_{data}^2) =0.995$ analysing the data
at $N_{side}=128$, and $P(\chi^2\le \chi_{data}^2) =0.994$ analysing
the data at $N_{side}=32,64,128$. These non-Gaussian deviations can be
clearly associated with the non-Gaussianity of the Archeops noise at
high resolution due to highly noisy pixels.
\subsection{Constraints on $f_{nl}$ for realistic non-Gaussian simulations}
\label{fnlconstraints}
The constraints in the $f_{nl}$ parameter were obtained using 300 CMB
non-Gaussian simulations such as the one described in
Sect.~\ref{model}.  and applying to the Archeops 143K03 bolometer data
the $\chi^2$ test in the case of non-Gaussian fluctuations defined by
an $f_{nl}$ parameter (Eq. \ref{chifnl}).  We computed an Archeops TOI
for each realization and from it we computed the corresponding Mirage
Archeops simulation for the $143K03$ bolometer. We did this for the
Gaussian and non-Gaussian part separately. So we have
\begin{equation}
d_{k03}(f_{nl}) = s_{g,k03}+f_{nl}*s_{ng,k03}+n_{g,k03}
\end{equation}
where $s_{g,K03}$ is a Gaussian CMB simulation, $s_{ng,K03}$ is its
corresponding non-Gaussian part, and $n_{g,K03}$ is a Gaussian
instrumental noise simulation.

We computed the mean value of the Minkowki functionals, $\langle V
\rangle_{f_{nl}}$, for $-1500 \leq f_{nl} \leq 1500$. We assumed that
in this interval the covariance matrix associated with them was
dominated by the Gaussian contribution: i.e. $C_{ij}(f_{nl}) \simeq
C_{ij}(f_{nl}=0)$.  Therefore, it was computed from $10^4$ Gaussian
simulations of signal and noise ($s_{g,k03}+n_{g,k03}$) of the
Archeops 143K03 bolometer. We obtained the $\chi^2(f_{nl})$ of
Archeops data given by Eqs. \ref{chifnl} and \ref{bigchifnl} for the
same combination of resolutions as the ones described in the above
Gaussianity analysis. In all cases we find the value of $f_{nl}$ that
minimizes $\chi^2(f_{nl})$. This is the best-fit value for the
$f_{nl}$ parameter. The significance of these values is estimated
using 2000 Gaussian simulations. For each simulation we computed
$\chi^2(f_{nl})$ vs $f_{nl}$ and obtained its best fit $f_{nl}$. At
the end we have a set of 2000 values of this parameter. As the
simulations are Gaussian, these values are centred around $f_{nl}=0$.

Table \ref{table_minfnl} lists the best-fit $f_{nl}$ to the data for
each case analysed. We also present the main properties of the
distribution of the $f_{nl}$ parameter as obtained from the
simulations.  The smaller dispersion corresponds to the case where we
combine the data at the three resolutions $N_{side}=$32, 64, and 128
(only 32\% of the available area).  Therefore this case leads to the
best constraints of $f_{nl}$ for the Archeops data. In
Fig. \ref{chi143K03fnl_ns32_ns64_ns128min90} we present the
$\chi^2(f_{nl})$ vs $f_{nl}$ of the Archeops data and the histogram of
the best fit $f_{nl}$ obtained from 2000 Gaussian simulations for this
optimal case.  From this we conclude that $f_{nl}= 70_{-400}^{+590}$
at 68\% CL and $f_{nl}= 70_{-920}^{+1075}$ at 95\% CL.

We may wonder if the constraints on the $f_{nl}$ parameter could be
improved by increasing the area at $N_{side}=128$. We have shown that,
if the noise were Gaussian, including the whole area available would
have produced very similar constraints. The reason is that the
excluded pixels were the noisiest ones and therefore increasing the
area would not improve the results.
\subsection{Comparing with the Sachs-Wolfe approximation} 
\begin{table*}[t]
  \center
  \caption{Best fit of Sachs-Wolfe $f_{nl}$ of Archeops data at
    different resolutions, mean, dispersion and some
    percentiles. \label{table_minswfnl}}
  \begin{tabular}{ccc|cc|cc|cc}
    \hline 
    \hline
    Resolution & Area (\%) & Best Fit $f_{nl}$  & $\langle f_{nl} \rangle$& $\sqrt{\langle f_{nl}^2 \rangle-\langle f_{nl} \rangle^2}$ & $X_{0.160}$& $X_{0.840}$ & $X_{0.025}$& $X_{0.975}$ \\ 
    \hline
    32 & 100 & -150 & 33 & 418 & -375 & 425 & -800 & 875 \\
    64 & 100 & 400 & -3 & 252 & -250 & 250 & -500 & 500 \\
    128 & 32 & -175 & 10 & 232 & -225 & 250 & -450 & 450  \\
    32,64 & 100,100 & 200 & 10 & 233 & -225 & 250 & -450 & 500 \\
    32,64,128 & 100,100,32 & 25 & 8 & 170 & -175 & 175 & -325 & 350\\
    \hline
    \hline
  \end{tabular}
\end{table*}
\begin{figure*}[t]
\center 
\epsfig{file=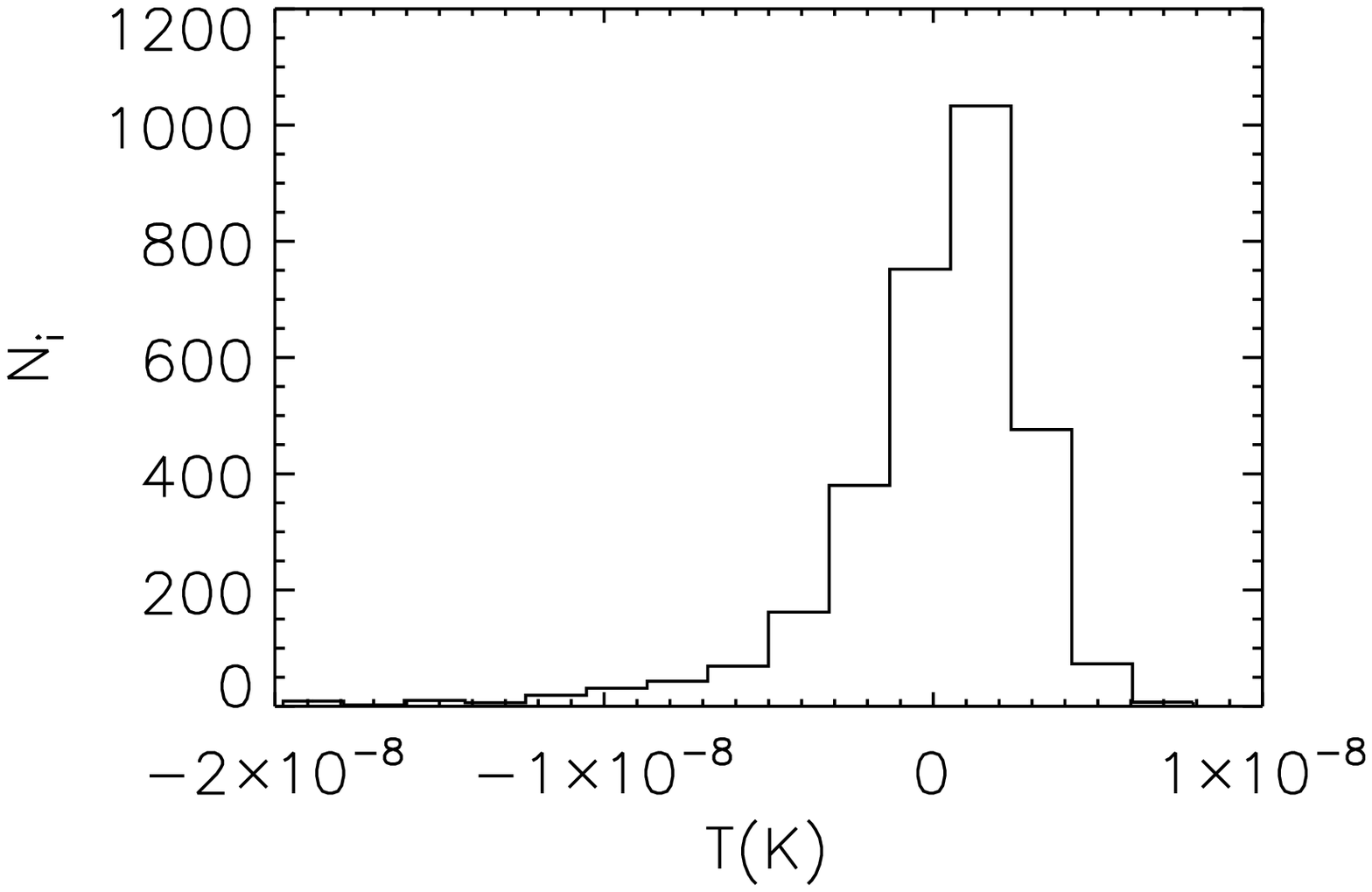,height=4.5cm,width=8cm}
\epsfig{file=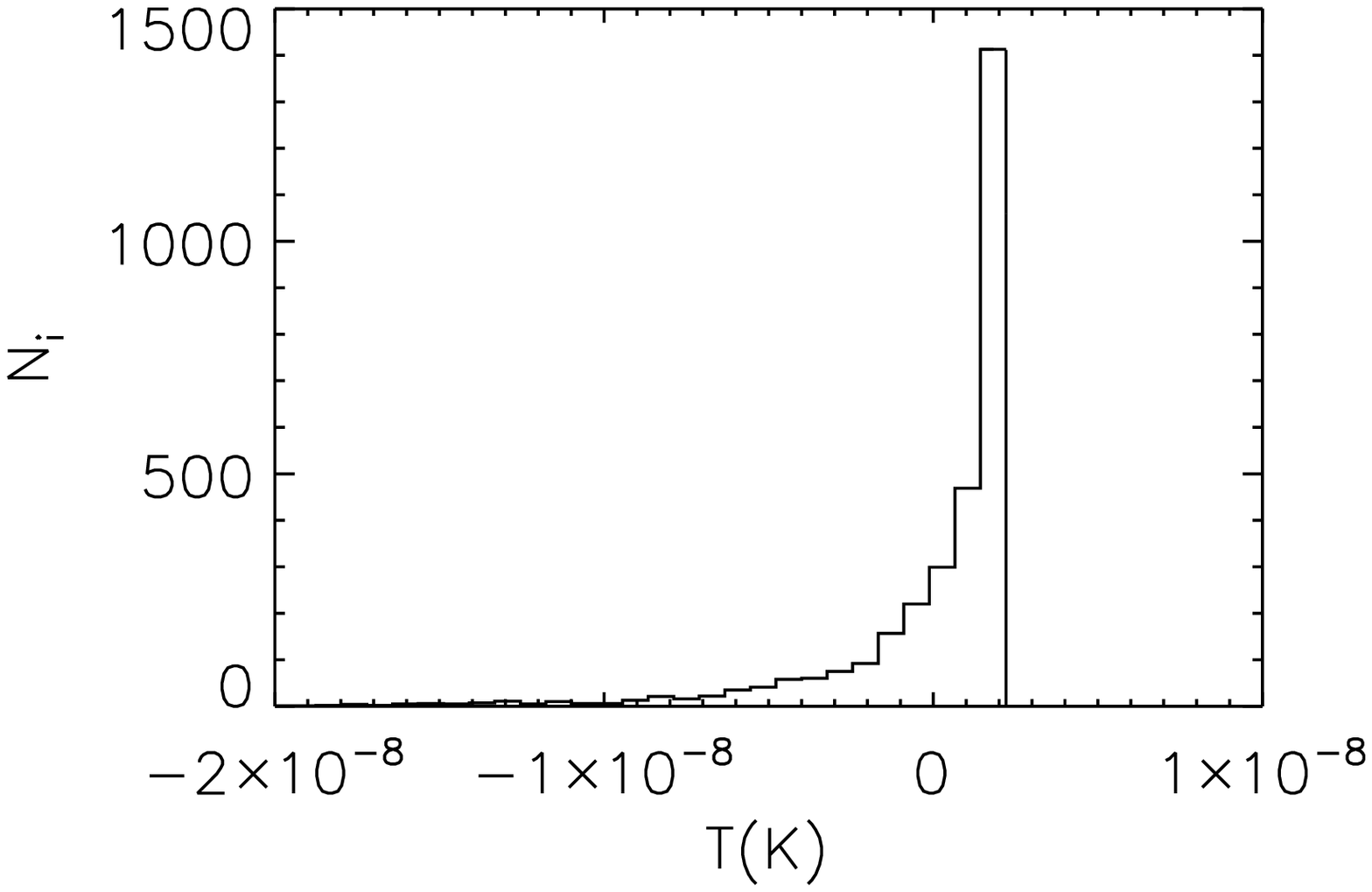,height=4.5cm,width=8cm}
\caption{{\it From left to right,} the non-Gaussian part of a full sky
CMB simulation at low resolution ($N_{side}=16$) and the corresponding
Sachs-Wolfe quadratic approximation.}
\label{comaprens16simus}
\end{figure*}
On large angular scales the main contribution to the CMB anisotropies
is given by the Sachs Wolfe effect: $\Delta T / T = -\phi / 3$
\citep{sachswolfe,komatsu2001}, where $\phi$ is the primordial
potential. Therefore, in this case, the temperature map can be
approximated in terms of a linear and a non-linear contribution in the
following simple way:
\begin{equation}
\frac{\Delta T}{T} = \frac{\Delta T_L}{T}-3f_{nl}\left(\frac{\Delta T_L^2}{T^2}-\left\langle\frac{\Delta T_L}{T} \right\rangle^2\right)
\end{equation}
where $\Delta T_L$ is Gaussian. We call this the Sachs-Wolfe
approximation and to avoid confusion hereafter the $f_{nl}$ parameter
in this case is called $f_{nl}^{SW}$. This approximation has been used
in several works \citep[for
example][]{cayon2003a,curto_32,smoot2007}. For this approximation the
Archeops non-Gaussian simulations can be obtained as follows
\begin{equation}
d_{k03}(f_{nl}) = s_{g,k03} - \frac{3f_{nl}^{SW}}{T_0}(s_{g,k03}^2 -
\langle s_{g,k03}^2\rangle) + n_{g,k03},
\end{equation}
where $s_{g,k03}$ and $n_{g,k03}$ are Gaussian CMB and noise
simulations respectively. As discussed above, this approximation is
only valid for angular resolutions of a few degrees
\citep{komatsu2001}. Nevertheless we performed the non-Gaussianity
analysis of the Archeops data at all considered scales in order to
constrain $f_{nl}^{SW}$ and then tested the validity of this
approximation as a function of the scale. In Table
\ref{table_minswfnl} we present the results of this analysis. We can
see that for all the considered resolutions, the $f_{nl}$ constraints
are better than for the case with the realistic simulations. This
difference becomes more important as the resolution increases. The
best constraints on $f_{nl}$ in this case are $f_{nl}^{SW}=
25_{-150}^{+200}$ at 68\% CL and $f_{nl}^{SW}=25_{-300}^{+375}$ at
95\% CL using the combination of the data at $N_{side}=$ 32, 64 , and
128 (only 32\% of the available area).

In general the Sachs-Wolfe approximation overestimates the
non-Gaussianity of the CMB fluctuations for a given $f_{nl}$
value. This can be clearly seen in Fig. \ref{comaprens16simus} where
we present from left to right the histogram of the non-Gaussian part
of a full sky CMB realistic simulation $\Delta T_{NG}$ and the
corresponding quadratic approximation $-3f_{nl}/T_0(\Delta
T_{G}^2-\langle \Delta T_{G} \rangle^2)$ at resolution $N_{side}=16$
i.e. a pixel size of 3.6 degrees. We can see that even at this scale
the simulations are very different. The approximation is just the
square of a Gaussian distribution centred to have zero mean whereas
the exact simulation is more Gaussian-like. This explains why the
constraints on the $f_{nl}$ parameters are tighter for the Sachs-Wolfe
approximation than for the exact case. Therefore, the results obtained
with the Sachs-Wolfe approximation should be taken with caution, since
the error bars are clearly underestimated.
%
%
\section{Conclusions}
In this paper we have presented a complete non-Gaussianity analysis of
the Archeops data at 143~GHz using the Minkowski functionals.  First,
we characterised the Archeops instrumental noise by taking the
difference of the data of the two most sensitive bolometers, 143K03
and 143K04. From this we found non-Gaussian deviations at high
resolution, 27 arcmin ($N_{side}=128$ in the Healpix pixelization
scheme). This is due to the noisiest pixels for which at high
resolution the number of observations per pixel does not allow good
systematic error removal. A more detailed analysis has been performed
for the $143K03$ bolometer for which a noise map was obtained by
subtracting the WMAP CMB data. From this analysis we found that pixels
with a number of observations below 90 were non-Gaussian.  Masking out
those pixels, the noise map is compatible with a Gaussian model.
Similar results were obtained with the $143K04$ bolometer although the
minimum number of observations per pixel for Gaussianity was of
the order of 150 and was not considered for further analysis.

Second, masking out these highly noisy pixels we performed a
Gaussianity analysis of the Archeops $143K03$ data at low and high
resolution. We found that the data are compatible with Gaussianity at
$N_{side}=32$, $N_{side}=64$, $N_{side}=128$, and for the combinations
$N_{side}=32,64$, and for the combinations $N_{side}=32,64,128$ at
better than 95\% CL.  From this analysis and using realistic
non-Gaussian simulations (\citet{liguori2003}) we imposed constraints
on the $f_{nl}$ parameter at these resolutions. The tightest
constraints are $f_{nl}= 70_{-400}^{+590}$ at 68\% CL and
$f_{nl}=70_{-920}^{+1075}$ at 95\% CL.

Third, we also imposed constraints on $f_{nl}$ using the Sachs-Wolfe
approximation, $f_{nl}^{SW}= 25_{-150}^{+200}$ at 68\% CL and
$f_{nl}^{SW}=25_{-300}^{+375}$ at 95\% CL. For comparison notice that
these constraints are a factor of $\approx3$ smaller than those given
in \citet{curto_32} where only low resolution ($N_{side}=32$) maps
were considered.

Finally, we performed a detailed comparison of the realistic
non-Gaussian simulations used in this paper and those from the
Sachs-Wolfe approximation. From this we conclude that even at low
resolution the Sachs-Wolfe approximation overestimates the
non-Gaussianity of the CMB fluctuations and therefore the $f_{nl}$
constraints imposed are too tight by a factor of three, as shown
above.
%
%
\begin{acknowledgements}
The authors thank the Archeops Collaboration for the possibility of
using Archeops data, M. Tristram for the Archeops simulation software,
and X. D\'esert for his useful comments. We also thank P. Vielva,
M. Cruz, C. Monteser\'in and J. M. Diego for useful discussions,
R. Marco and L. Cabellos for computational support. We acknowledge
partial financial support from the Spanish Ministerio de Educaci\'on y
Ciencia (MEC) project AYA2007-68058-C03-02. A.C. thanks the Spanish
Ministerio de Educaci\'on y Ciencia (MEC) for a pre-doctoral FPI
fellowship.  S.M. thanks ASI contract I/016/07/0 "COFIS" and ASI
contract Planck LFI activity of Phase E2 for partial financial
support. The authors acknowledge the computer resources, technical
expertise and assistance provided by the Spanish Supercomputing
Network (RES) node at Universidad de Cantabria. We acknowledge the use
of Legacy Archive for Microwave Background Data Analysis
(LAMBDA). Support for it is provided by the NASA Office of Space
Science. The HEALPix package was used throughout the data analysis
\citep{healpix}.
\end{acknowledgements}
\end{document}